\documentclass[twocolumn,showpacs,preprintnumbers,superscriptaddress,aps,prd,nofootinbib]{revtex4-2}

\usepackage[dvipdfmx]{graphicx}
\usepackage{dcolumn}
\usepackage{bm}
\usepackage{amsmath,amssymb,amsfonts}
\usepackage{latexsym}
\usepackage[flushleft]{threeparttable}
\usepackage{bbm}
\usepackage{color}
\usepackage{hyperref}
\usepackage{cancel}
\usepackage[english]{babel}

\hypersetup{
  colorlinks  = true,
  urlcolor    = blue,
  linkcolor   = red,
  citecolor   = blue
}

\usepackage[utf8]{inputenc}
\usepackage[normalem]{ulem}

\newcommand{\cmmnt}[1]{}
\usepackage{soul}

\newcommand\yd[1]{{\color{blue} #1}}

\newcommand{\eqal}[2]{\begin{equation}\begin{aligned}  #2 \end{aligned}\end{equation}}

\begin{document}

\preprint{ACFI-T23-04}

\title{Phase transitions, anomalous baryon number violation and electroweak multiplet dark matter
} 

%
\author{Yanda Wu}
\email{yanda.wu7@sjtu.edu.cn}
\affiliation{Tsung-Dao Lee Institute, Shanghai Jiao Tong University, Shanghai 201210, China}
\affiliation{Shanghai Key Laboratory for Particle Physics and Cosmology, 
Key Laboratory for Particle Astrophysics and Cosmology (MOE), 
Shanghai Jiao Tong University, Shanghai 200240, China}
\author{Wenxing Zhang}%
\email{zhangwenxing@hbu.edu.cn}
\affiliation{Department of Physics, Hebei University, Baoding, 071002, China}
\affiliation{Hebei Key Laboratory of High-Precision Computation and Application of Quantum Field Theory, Baoding 071002, China}
\affiliation{Hebei Research Center of the Basic Discipline for Computational Physics, Baoding 071002, China}
\author{Michael J.~Ramsey-Musolf}%
\email{mjrm@sjtu.edu.cn, mjrm@physics.umass.edu}
\affiliation{Tsung-Dao Lee Institute, Shanghai Jiao Tong University, Shanghai 201210, China}
\affiliation{Shanghai Key Laboratory for Particle Physics and Cosmology, 
Key Laboratory for Particle Astrophysics and Cosmology (MOE), 
Shanghai Jiao Tong University, Shanghai 200240, China}
\affiliation{Amherst Center for Fundamental Interactions, Department of Physics,
University of Massachusetts Amherst, MA 01003, USA }
\affiliation{ Kellogg Radiation Laboratory, California Institute of Technology,
Pasadena, CA 91125, USA}

\bigskip

\date{\today}

\begin{abstract}
We perform a comprehensive analysis of baryon number violation during an electroweak phase transition (EWPT) within the framework of a scalar electroweak multiplet extension of the Standard Model. We classify the multiplet representations, topological properties, and corresponding thermal histories. Sphaleron or monopole topological field solutions emerge during the EWPT depending on the stage of the phase transition and the hypercharge of the new scalar multiplet. Furthermore, the monopole field solution pertains when the neutral component of the additional scalar multiplet is a viable dark matter candidate.
We further analyze other formal considerations, including the construction of the \lq\lq sphaleron matrix\rq\rq\, for higher dimensional representations, computation of the sphaleron and monopole masses, and the choice of boundary conditions when solving the field equations of motion. We apply these considerations to the computation of sphaleron energy and monopole mass within the context of a multi-step EWPT, employing the SU(2)$_L$ septuplet scalar extension to the Standard Model (SM) as a case of study from the minimal dark matter paradigm. For the first step of a two-step EWPT, we delineate the relationship between the monopole mass and the parameters relevant to dark matter phenomenology.
\end{abstract}


\maketitle

\section{Introduction}
The origin of baryon asymmetry of the universe (BAU) remains an open question at the frontier of particle physics and cosmology. In order to explain the BAU, Sakharov proposes three necessary conditions: (1) baryon-number violation; (2) C and CP violation; (3) departure from thermal equilibrium or CPT violation \cite{Sakharov:1967dj}. 
In principle, the Standard Model (SM) provides all the necessary ingredients for generation of the baryon asymmetry during the era of electroweak symmetry-breaking (EWSB), a scenario known as electroweak baryogenesis (EWBG).
Indeed, the first condition can be fulfilled by the non-perturbative weak sphaleron process. However, the SM fails to satisfy the second and third conditions. The CP-violation associated with the Cabibbo-Kobayashi-Maskawa matrix is too weak to generate the observed BAU \cite{Gavela:1993ts,Huet:1994jb,Gavela:1994dt}, and EWSB occurs through a smooth crossover transition due to the large Higgs mass \cite{Kajantie:1996mn,Gurtler:1997hr,Laine:1998jb,Csikor:1998eu,Aoki:1999fi}, thereby missing the needed out of equilibrium requirement. Many beyond Standard Model (BSM) theories have been proposed to remedy these shortcomings and facilitate EWBG (see Refs.~\cite{Morrissey:2012db,Ramsey-Musolf:2019lsf,Bodeker:2020ghk} for reviews). In this work, we focus on a key element of BSM EWBG: electroweak sphaleron and monopole dynamics. We do so in the context of a general class of BSM scenarios, namely, those involving an extended Higgs sector containing higher dimensional electroweak multiplets. 

Electroweak baryogenesis requires a first order electroweak phase transition (FOEWPT), during which bubbles of broken symmetry nucleate in the symmetric phase. The BSM CP-violating interactions at the bubble walls generate a left-handed fermion number density that biases symmetric phase sphaleron transitions into generation of non-zero baryon plus lepton number ($B+L$) \cite{Kuzmin:1985mm, Rubakov:1996vz,Morrissey:2012db}. The asymmetry diffuses into the bubble interiors. A sufficiently \lq\lq strong\rq\rq\ FOEWPT leads to suppression of the broken phase sphaleron rate, thereby allowing preservation of the asymmetry \cite{Patel:2011th}. A central question, therefore, pertains to the broken phase sphaleron rate: is it sufficiently quenched so as to preserve the BAU? 

While the most reliable approaches to answering this question are obtained using lattice computations, as a practical matter performing a broad survey of BSM scenarios and associated parameter choices relies on (semi-)analytic methods and perturbation theory. The latter provides a baseline for comparison and validation against non-perturbative studies. The aim of the following study is to refine this baseline and clarify some formal considerations along the way. In doing so, we recall that the analytic result for the broken phase sphaleron rate, $\Gamma_\mathrm{sph}$ can be written as the product of a dynamical prefactor $A$ and a statistical factor \cite{Arnold:1987mh, Carson:1990jm, Baacke:1993aj}:
\begin{equation}
\Gamma_\mathrm{sph} = A\,  e^{-E_\mathrm{sph}/T}\ \ \ ,
\end{equation}
where $E_\mathrm{sph}$ is the energy associated with the semiclassical sphaleron solution.
Our focus in the present study falls on the latter. 

\begin{figure*}[t]
\center
\includegraphics[width=14cm]{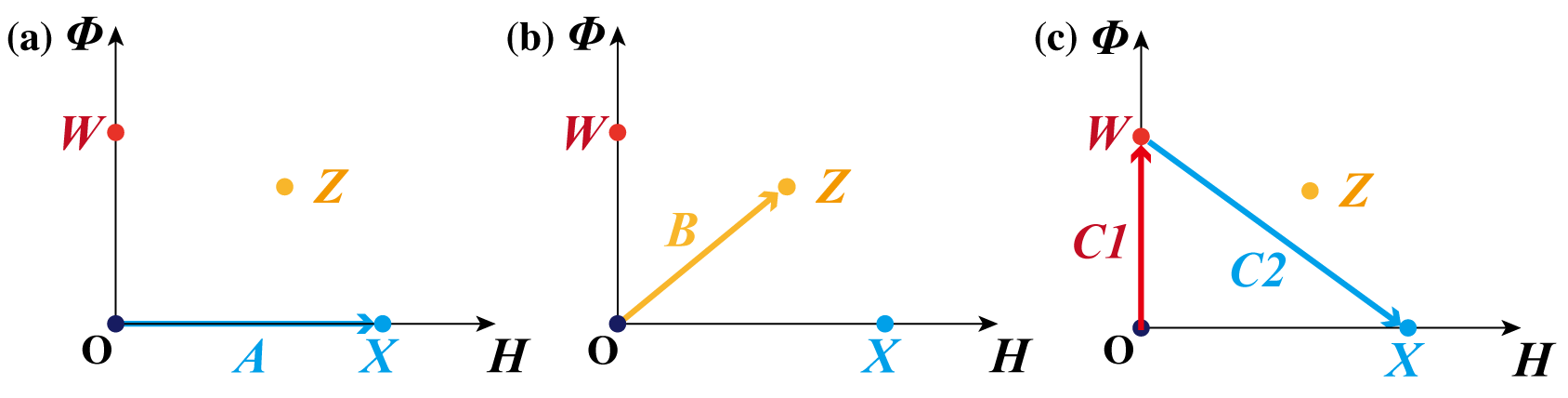}
\caption{Three possible EWSB patterns. In each plot, the horizontal axis represents the SM Higgs vacuum, the vertical axis stands for the vacuum of BSM (electroweak multiplet in our study) particle. The $O$ denotes the symmetric phase, while $X, W$ and $Z$ represent the three stationary points of the total potential, i.e. the solutions for which the first derivative of the potential with respect to the fields vanishes. (a) represents a one-step EWPT from symmetric phase to pure Higgs phase; (b) denotes a one-step EWPT to the mixed phase where the Higgs particle and scalar multiplet particle both obtain a vacuum expectation value (vev); (c) depicts a two-step EWPT scenario, where the first step moves from symmetric phase to the pure multiplet phase, and the second step moves from the multiplet phase to the Higgs phase. 
}
\label{fig:different_steps_EWPT}
\end{figure*}

To further set the context, we recall that the thermal history of EWSB can entail either a single, direct transition to the present \lq\lq Higgs phase\rq\rq\,  or a series of steps. In the presence of additional scalar fields $\Phi$, a different vacuum associated with a non-zero vacuum expectation value (vev) for one or more components of $\Phi$ may precede the Higgs phase. Alternately, the Higgs phase may also involve a non-zero $\Phi$ vev. While $\Phi$ may be either a SM gauge singlet or carry SM quantum numbers, in this study we consider the case where $\Phi$ is an SU(3$)_C$ singlet but charged under SU(2$)_L\times$U(1$)_Y$. We further specify  that only the neutral component of $\Phi$ obtains a non-zero vev. Three representative patterns of EWSB are illustrated in Fig.~\ref{fig:different_steps_EWPT}, where case (a), (b) and (c) represent the SM one-step EWPT, one-step EWPT to the mixed phase and two-step EWPT, respectively. 

Each case may accommodate EWBG. For the single-step transitions in (a) and (b), the presence of $\Phi$ will modify $\Gamma_\mathrm{sph}$ through thermal loops and, for (b), through additional contributions to the semiclassical sphaleron solution. Note that for (b), constraints from the electroweak $\rho$-parameter place strong constraints on $\langle\Phi \rangle$, when $\Phi$ is neither a gauge singlet or second Higgs doublet. One may evade these constraints through a suitable choice of field content, as in the Georgi-Machacek model \yd{\cite{Georgi:1985nv}}. For (c), the first step may accommodate EWBG if (i) this step involves a FOEWPT; (ii) if BSM CP violation (CPV) interactions generate a sufficiently large asymmetry; (iii)  the rate of baryon number violating (BNV) processes in the EWSB $\Phi$ vacuum is sufficiently suppressed; and (iv) the second step to the Higgs phase does not allow for re-excitation of the EW sphalerons. The viability of this possibility has been demonstrated in Refs.~ \cite{Patel:2012pi,Blinov:2015sna,Niemi:2020hto}.
To our knowledge, BNV dynamics for these scenarios in the presence of $\Phi$ have not been explored in a unified and systematic way. In what follows we endeavor to do so, focusing on cases (b) and (c) wherein $\Phi$ can play an active role in the semiclassical solutions to the field equations associated with BNV. We investigate both the corresponding topological structure and relevant energy scale.

To that end, this study mainly consists of two parts. In the first part, we review, update, and clarify various formal aspects related to the semiclassical treatment of the BNV solutions in BSM theories,  
including: topological properties associated with different representations; relationships between various treatments for the BNV configurations; a general construction of the 1-form framework for a general scalar multiplet;  and choice of boundary conditions when solving equation of motion. Most importantly, we show that when  the $\Phi$-vev alone breaks electroweak symmetry, the corresponding BNV configuration is  the sphaleron for $\Phi$ having non-zero hypercharge, $Y$, or a monopole for $Y(\Phi)=0$. Dark matter direct detection constraints imply that only a $Y=0$ multiplet with vanishing vev (point $X$ in Fig.~\ref{fig:different_steps_EWPT}) can be a viable dark matter candidate \cite{Bottaro:2022one,Panci:2024oqc}. Thus, phase $W$ in the two-step scenario (c) would be associated with monopole-catalyzed BNV processes.

In the latter part, we compute the sphaleron energy for scenarios (b) and (c) for $\Phi$ being an electroweak septuplet, whose presence in the Higgs vacuum of scenario (c) can contribute to the dark matter (DM) relic density. In this instance, we delineate the dependence of the monopole mass on the parameters relevant to DM phenomenology: the DM mass, its self-interaction, and the coupling to SM fields that enters the annihilation and direct detection cross sections. In our current work, we primarily focus on the analysis of the zero-temperature model. Our aim is to provide a methodology for applying the sphaleron and monopole formalism to different EWSB patterns, where the zero-temperature model can provide a good approximation of physical quantities. The thermally corrected model can be analyzed in a parallel manner. We find that, depending on the values of these parameters, the monopole mass for step C1 of case (c) can be significantly larger than in the SM electroweak sphaleron energy for single step transition in case (a), suggesting that the two-step scenario can be particularly conducive to EWBG. 

Our discussion of these issues is organized as follows. In Sec.~\ref{sec::theory_formalism}, we present a detailed analysis of the theory formalism, applicable to both the SM and BSM scenarios. In Sec.~\ref{sec:EWPT_model_analysis}, we discuss an electroweak multiplet extension to the SM and present two possible types of EWPT involving this extension. In Sec.~\ref{sec:sphaleron_energy}, we compute the corresponding sphaleron energy and monopole mass for this model as appropriate to the EWSB scenario. Our conclusions are summarized in Sec.~\ref{sec:conclusion}. In the Appendix, we present other sphaleron configurations, computation details of sphaleron energy, monopole mass and model mass matrices.

\section{Theory formalism} \label{sec::theory_formalism}
In this section, we demonstrate the properties of topological field solutions when the SM Higgs ($H$) is extended by a general-dimensional SU(2) multiplet $\Phi$. As depicted in Fig.\ref{fig:different_steps_EWPT}, there are three notable scenarios: the SM Higgs acquiring a vev, the multiplet acquiring a vev, and both acquiring vevs simultaneously, corresponding to vacuum $X$, $W$, and $Z$, respectively. Below are the key features demonstrated in this section:
\begin{itemize}
    \item The classification of topological field solutions when an SU(2) multiplet $\Phi$ is added to the SM field content. If $\Phi$ acquires a vev as shown at point $W$ in Fig.\ref{fig:different_steps_EWPT}, a monopole or sphaleron solution emerges depending on whether its hypercharge is zero or not. When the SM Higgs and $\Phi$ acquire vevs simultaneously, the field solution is a sphaleron. 
    \item We construct the general-dimensional sphaleron and monopole matrices for the first time. The sphaleron matrix $U_{\text{sph}}$ is constructed from the multiplication of two Wigner matrices. Later, we demonstrate that the monopole matrix can be derived from the sphaleron matrix when the non-contractible loop parameter $\mu$ in $U_{\text{sph}}$ equals $1/2$.
    \item We provide a comprehensive comparison of the mechanisms for sphaleron and monopole-induced baryon number violation (BNV). 
    Although the mechanism for monopole-induced BNV has received less attention in EWBG discussions compared to the corresponding sphaleron-induced BNV, it is nevertheless relevant for the scenarios considered in this work. 
    \item We present a derivation of the sphaleron energy and monopole mass for a general electroweak multiplet and clarify the requirements on the corresponding boundary conditions used when obtaining the field equation solutions.
\end{itemize}
We also address other formal aspects in this section: a summary of the main choices of different sphaleron configurations and their relationships, and the computation of the $1/2$ baryonic charge of the sphaleron configuration. 

\subsection{Topological field solutions for a general-dimensional SU(2) multiplet}
\label{sec:classification_of_topological_field_solutions}
We first discuss the situation when a single multiplet acquires a vev after EWSB. The classification of field solutions is related to the mapping of static space at infinity (which is a 2-sphere, $S^2$) to the vacuum manifold of the theory. The vacuum manifold, denoted by $\mathcal{M}$, is characterized by the coset space \cite{Preskill:1984gd}:
\begin{equation} \label{eq:cosetspace}
\mathcal{M}=\frac{G}{H}=\frac{SU(2)_L\times U(1)_Y}{H}.
\end{equation}
where $G$ and $H$ are the symmetry groups before and after EWSB, respectively. 
The residual symmetry group is related to the representation of the multiplet, specifically its hypercharge, $Y$. For a  
$Y=0$ multiplet, 
$H=U(1)_Y\times U(1)_{\text{em}}$, since the multiplet electric charge is the eigenvalue solely of $T_3$.
For a  
$Y\not=0$ multiplet, $H=U(1)_{\text{em}}$, with generator $Q=T_3+Y$, is the only remaining symmetry group.
In the foregoing cases, the number of unbroken generators after EWSB are two and one, respectively. 

For a  $Y\not=0$ multiplet, the resulting topological field solution is the sphaleron \cite{Manton:1983nd}:
\begin{align}
\mathcal{M}&=\frac{SU(2)_L\times U(1)_Y}{U(1)_{\text{em}}}\sim S^3,\ \ \pi_3(\mathcal{M})=Z,
\end{align}
where the symbol $\sim$ denotes an isomporphism and  $\pi_3(\mathcal{M})$ denotes the mapping of a 3-sphere to $\mathcal{M}$ at spatial infinity. Such a 3-sphere is formed by the non-contractible loop parameter $\mu$ and the spatial angular parameters $\theta$ and $\phi$. Without the dependence on $\mu$, the mapping would be $\pi_2(S^3)$, which is the null mapping associated with continuous deformation to the vacuum configuration. Thus, the existence of the solution requires the existence of an additional parameter, $\mu$ to yield $\pi_3(S^3)$. As we will see when analyzing the impact of the sphaleron on topological charge, $\mu$ will play the role of time. For a given, fixed choice of $\mu$, the sphaleron configuration can be continuously deformed into the vacuum configuration, since $\pi_2(S^3)=0$, leading to the unstable nature of sphaleron. 

Under these considerations, one may also write the sphaleron scalar and gauge field configuration, following the convention by \cite{Manton:1983nd}. 
\begin{align}
\label{eq:sphaleron_multiplet_confg_manton}
\Phi &= \frac{v_\phi}{\sqrt{2}}\phi(\xi)U_{\text{sph}}(\mu,\theta,\phi).\begin{pmatrix}
    0\\
    \cdots\\
    1 \\
    \cdots \\
    0
\end{pmatrix},\\ 
\label{eq:sphaleron_gauge_confg_manton}
A_i^a T^a dx^i &= -\frac{i}{g}f(\xi) (\partial_i U_{\text{sph}} )U_{\text{sph}}^{-1}dx^i,
\end{align}
Here, $v_\phi$ denotes the vev of the multiplet, while $\phi(\xi)$ and $f(\xi)$ are radial profile functions for the scalar and gauge field, with $\xi\equiv gv_\phi r$ being the dimensionless radial parameter. In the gauge field configuration, the $i$ index in $A_i^a$, $\partial_i$, and $dx^i$ is contracted, with $i\in[\theta,\phi]$ and $dx^i\in[d\theta,d\phi]$. The construction of $U_{\text{sph}}$ for an arbitary SU(2) representation $J$ entails substantial technique details, which we defer to Eq.~(\ref{eq::general_u}). For clarity, we first present the fundamental representation of $U_\text{sph}$:\footnote{Strictly speaking, the sphaleron configuration is defined only at $\mu=\pi/2$. For other values of $\mu$, $U_\text{sph}$ gives a field ansatz for $A_i^a$.}
\begin{equation}
\label{eq::Manton_Umatrix_fundamental}
\begin{aligned}
    & U_{\text{sph}}(\mu,\theta,\phi)= \\
    &\left [\begin{array}{cc}
			e^{i \mu } (\cos \mu -i \cos \theta  \sin \mu ) & e^{i \varphi } \sin \theta  \sin \mu            \\
			-e^{-i \varphi } \sin \theta  \sin \mu          & e^{-i \mu } (\cos \mu +i \cos \theta  \sin \mu) \\
		\end{array}\right].   
\end{aligned}
\end{equation}

For a $Y=0$  multiplet, the resulting topological field solution is a monopole \cite{Preskill:1984gd}:
\begin{align} \label{eq:topology_for_monopole}
\mathcal{M}&=\frac{SU(2)_L\times U(1)_Y}{U(1)_{\text{em}}\times U(1)_Y}\sim S^2,\ \pi_2(\mathcal{M})=Z,
\end{align}
with corresponding field profiles
\begin{align}
\label{eq:mon_multiplet_manton}
\Phi &= \frac{v_\phi}{\sqrt{2}}\phi(\xi)U_{\text{mon}}(\theta,\phi).\begin{pmatrix}
    0\\
    \cdots\\
    1 \\
    \cdots \\
    0
\end{pmatrix},\\
\label{eq:mon_gauge_manton}
A_i^a T^a dx^i &= -\frac{i}{g}f(\xi) (\partial_i U_{\text{mon}} )U_{\text{mon}}^{-1}dx^i,
\end{align} where $\pi_2(\mathcal{M})$ denotes the mapping of a 2-sphere to $\mathcal{M}$ at spatial infinity. Such mapping is stable, implying that the monopole is stable. In contrast to the $Y\not=0$ case, stability of the mapping requires no additional dependence on a variable that might play the role of time.  Note that in Eq.~(\ref{eq:topology_for_monopole}) the $U(1)_Y$ factor is redundant since gauge invariance precludes renormalizable couplings of $\Phi$ to the fermion sector. Nevertheless, we include it here for pedagogical clarity.

Consider, now, the form of the \lq\lq monopole matrix\rq\rq ,\,$U_{\text{mon}}$, and its relationship to $U_{\text{sph}}$.
To that end, first note that $U_{\text{mon}}$ depends on only the two spatial angular co-ordinates,
$\theta$ and $\phi$. 
Second, for the specific case when $\Phi$ is a real triplet (adjoint representation), we expect the monopole gauge field configuration to coincide with the 't Hooft-Polyakov monopole solution \cite{tHooft:1974kcl,Polyakov:1974ek}:
\begin{align} \label{eq:Hooft_monopole_gauge_confg}
A_0=0,\quad A_i^a=c_1 \epsilon_{iab} x^b \frac{F(\xi)}{gr^2},
\end{align}
where $A_0$ is the temporal gauge field, $A_i^a$ is the spatial gauge field with spatial index $i$ and isospin index $a$; $c_1$ is a normalization constant and $F(\xi)$ is a dimensionless radial function.
Below, we show explicitly how Eq.~(\ref{eq:Hooft_monopole_gauge_confg}) follows from Eq.~(\ref{eq:mon_gauge_manton}) for any $Y=0$ scalar multiplet representation.

We will shortly show how to derive $U_{\text{mon}}$ starting from $U_{\text{sph}}$. In doing so, it is also helpful to review the various equivalent choices of the sphaleron field configuration that have appeared in the literature. 
The configuration given in Eqs.~(\ref{eq:sphaleron_multiplet_confg_manton},\ref{eq:sphaleron_gauge_confg_manton}) 
follows Manton and Klinkhamer's convention, which we denote as the MK configuration. There is another widely used sphaleron configuration, proposed by Klinkhamer and Laterveer \cite{Klinkhamer:1990fi}, which we denote as the KL configuration.
We now show the equivalence of this KL field configuration with the MK configuration.
Applying a unitary transformation $(U_{\text{sph}})^{-1}$ to the MK configuration in Eq.~\eqref{eq:sphaleron_multiplet_confg_manton}, the  multiplet field becomes:
\begin{equation} \label{eq::gauge_trans_Higgs}
    \Phi \rightarrow \frac{v_\phi}{\sqrt{2}}\phi(\xi)\begin{pmatrix}
    0\\
    \cdots\\
    1 \\
    \cdots \\
    0
\end{pmatrix} \ \ \ .
\end{equation}
Similarly, applying a unitary transformation $(U_{\text{mon}})^{-1}$ to the monopole multiplet configuration, Eq.~(\ref{eq:mon_multiplet_manton})
, yields the identical form the multiplet configuration as in Eq.~\eqref{eq::gauge_trans_Higgs}.
In both cases, the gauge field transforms in the usual way: $A_\mu^a T^a\rightarrow U A_\mu^a T^a U^{-1}-\frac{i}{g}(\partial_{\mu}U) U^{-1}$, implying that
\begin{equation}
\begin{aligned} \label{eq::gauge_trans_Gauge}
A_i^a T^a&\rightarrow U^{-1}\left(-\frac{i}{g}f(\xi)(\partial_{i}U) U^{-1}\right)U  -\frac{i}{g}(\partial_{i}U^{-1})U \\
&=\frac{i}{g}(1-f(\xi))U^{-1} \partial_{i}U\ \ \ , \\
\end{aligned}
\end{equation}
where $U$ denotes either $U_{\text{sph}}$ or $U_{\text{mon}}$.

The forms for the scalar multiplet and gauge field configurations in Eqs.~(\ref{eq::gauge_trans_Higgs},\ref{eq::gauge_trans_Gauge}) are the same as the phase II KL configuration under the constraint that $f_0=0$ and $f_3=f$, where $f_0$ and $f_3$ are the U(1) field and $A_i^3$ field radial profile functions. We provide the full expressions of KL configurations and definitions of the three phases of NCL in later Sec.~\ref{sec:Sphaleron energy, monopole mass and equation of motions}.

Despite this similarity of forms, the full parameter dependencies of $U_{\text{sph}}$ and $U_{\text{mon}}$ differ significantly, as dictated by the aforementioned topological considerations.
The sphaleron matrix is parameterized by spherical angular parameters $\theta$ , $\phi$ and the NCL parameter $\mu$, while the monopole matrix carries no $\mu$-dependence. For each case we may write
\begin{align} \label{eq:one_form_Fa_definition}
iU^{-1} \partial_i U dx^i = \sum_{a=1}^3 F_a T^a,
\end{align}
where $T^a$ are SU(2) generators and $F_a$ are scalar functions of the relevant parameters.

Focusing now on the sphaleron case, $U=U_{\text{sph}}$, for $\Phi$ in the fundamental representation the $F_a$ are given as\footnote{Analogous to Eq.~(\ref{eq::Manton_Umatrix_fundamental}), the sphaleron configuration is defined only at $\mu=\pi/2$.}
\begin{equation}
	\begin{aligned} \label{eq:one_form_definition}
		&F_1=-(2\sin^2\mu\cos(\mu-\phi)-\sin2\mu\cos\theta\sin(\mu-\phi))d\theta \\
		& -(\sin2\mu\cos(\mu-\phi)\sin\theta+\sin^2\mu\sin2\theta\sin(\mu-\phi))d\phi,\\
		&F_2=-(2\sin^2\mu\sin(\mu-\phi)+\sin2\mu\cos\theta\cos(\mu-\phi))d\theta \\
		& +(-\sin2\mu\sin(\mu-\phi)\sin\theta+\sin^2\mu\sin2\theta\cos(\mu-\phi))d\phi,\\
		&F_3=-\sin2\mu\sin\theta d\theta+2\sin^2\theta \sin^2\mu d\phi.
	\end{aligned}
\end{equation}
It is useful to provide a general construction of
$U_{\text{sph}}$ and the 1-form $F_a$ applicable to a general scalar SU(2$)_L$ multiplet of arbitrary isospin $J$.\footnote{In passing, we note that
Ahriche et al. \cite{Ahriche:2014jna} calculate the sphaleron energy for higher dimensional $\text{SU}(2)$ scalar representations, wherein they use but do not prove that the 1-form $F_a$ is invariant concerning different representation dimensions. We will expand on their work by showing this invariance.}
To that end, it is natural to consider the parametrization of an arbitrary SU(2) matrix in terms of the Wigner-D matrix. We consider the general sphaleron matrix:
\begin{align} \label{eq:U_sph_general_single_wigner_D_attempt}
U_\text{sph}(\alpha,\beta,\gamma)=\mathcal{D}^{1/2}\big(\alpha,\beta,\gamma \big)=e^{-i\alpha T^3} e^{-i\beta T^2} e^{-i\gamma T^3},
\end{align}
where the Euler angles $\alpha,\beta,\gamma$ are representation-independent. To determine their value, we first express this in the fundamental representation:
\begin{align}\label{eq::su2_2dim_wigner} \nonumber
&\mathcal{D}^{1/2}\big(\alpha,\beta,\gamma \big)  \\ \nonumber
    &=e^{-i\alpha \frac{\sigma_3}{2}} e^{-i\beta \frac{\sigma_2}{2}} e^{-i\gamma \frac{\sigma_3}{2}}\\
	&=\left(
	\begin{matrix}
		e^{-i\frac{\alpha + \gamma}{2}}\cos\frac{\beta}{2}  & -e^{-i\frac{\alpha - \gamma}{2}} \sin\frac{\beta}{2}  \\
		e^{i\frac{\alpha - \gamma}{2}} \sin\frac{\theta}{2}  & e^{i\frac{\alpha + \gamma}{2}} \cos\frac{\beta}{2} 
	\end{matrix}
	\right).
\end{align}
We then equate this matrix with the sphaleron matrix in Eq.~\eqref{eq::Manton_Umatrix_fundamental}, we can obtain the following relationships:
\begin{equation}
\begin{aligned} \label{eula_parameter}
&\cos \left(\frac{\beta }{2}\right) \cos \left(\frac{\alpha }{2}+\frac{\gamma }{2} \right)=1+\sin^{2}\mu(\cos\theta-1), \\
&\sin \left(\frac{\beta }{2}\right) \sin \left(\frac{\alpha }{2}-\frac{\gamma }{2}\right)=\sin\phi \sin\theta \sin\mu, \\
&\sin \left(\frac{\beta }{2}\right) \cos \left(\frac{\alpha }{2}-\frac{\gamma }{2}\right)=-\cos\phi \sin\theta \sin\mu , \\
&\cos \left(\frac{\beta }{2}\right) \sin \left(\frac{\alpha }{2}+\frac{\gamma }{2}\right)=\sin\mu \cos\mu (\cos\theta-1)\ \ .
\end{aligned}
\end{equation}
We obtain these relations by first expanding Eqs.~\eqref{eq::Manton_Umatrix_fundamental} and \eqref{eq::su2_2dim_wigner} as a linear combination of the basis matrices $I_{2\times 2}, \sigma_1, \sigma_2$ and $\sigma_3$; then equating the corresponding expansion coefficients. 
While it is possible in principle to solve these equations and establish relationships between $(\alpha$,$\beta$,$\gamma$) and ($\mu$,$\theta$,$\phi$), doing so in practice is cumbersome. Not only must we be careful with the sign of the final solution of three Euler angles, but also they have non-linear dependence with $\mu,\theta,\phi$, complicating the calculation of the $F_a$ in Eq.~(\ref{eq:one_form_Fa_definition}). Therefore, although Eq.~\eqref{eq:U_sph_general_single_wigner_D_attempt} looks quite intuitive, we seek an alternate method. 

Instead, we can use the multiplication of multiple Wigner-D matrices to represent the sphaleron matrix. For a general representation $J$ with matrix dimension $2J+1$, we can write the sphaleron matrix as
\begin{equation}
   \begin{aligned}
    \label{eq::general_u} 
	 &U_{\text{sph}}\big(\mu, \theta, \phi\big) \\
     &= D^{J}\big(\omega_{-}, -\theta, \mu \big) D^{J}\big(\mu, \theta, \omega_+\big)\\
     &=e^{-i\omega_-T^3}e^{i\theta T^2}e^{-2i\mu T^3}e^{-i\theta T^2} e^{-i\omega_+ T^3}, 
\end{aligned} 
\end{equation}
with
\begin{equation}
    \omega_{\pm}=-\mu\pm(\phi-\frac{\pi}{2})\ \ \ .
\end{equation}
if we set $J=1/2$, we can restore the standard sphaleron matrix Eq.~\eqref{eq::Manton_Umatrix_fundamental}. The higher-dimensional form of $U_\text{sph}$ then follows directly from Eq.~(\ref{eq::general_u}) by replacing the SU(2) generators $T^i$ with their higher-dimensional counterparts. For example, for $J=1$, one obtains $U_\text{sph}$ by substituting the following expressions for the generators $T^2$ and $T^3$:
\begin{align} \label{eq:T2_T3_generators_J_1} T^2=\frac{i}{\sqrt{2}}\begin{pmatrix}
    0 & -1 & 0 \\
    1 & 0 & -1 \\
    0 & 1 & 0
\end{pmatrix},\quad T^3=\begin{pmatrix}
    1 & 0 & 0 \\
    0 & 0 & 0 \\
    0 & 0 & -1
\end{pmatrix}. 
\end{align}
With this kind of parametrization method it is relatively straightforward to calculate the 1-from, since the Euler parameters are linearly related to $\mu, \theta$ and $\phi$. 
The 1-form $F_a$ can then be calculated through
\begin{equation}
	F_a = \left[\text{Tr}(T^3)^2\right]^{-1} \text{Tr}[iU_{\text{sph}}^{-1}\partial_iU_{\text{sph}}.T^a]dx^i.
\end{equation}
Using this calculation method, we have verified explicitly that one obtains the same $F_a$ for $J=[1,3/2,2,5/2,3]$ under the $\text{SU}(2)$ generator representation \cite{GroupMath}. 

Now consider the construction of the monopole matrix for a general SU(2) multiplet. We start by multiplying the sphaleron matrix with $\mu=\frac{\pi}{2}$ by a ($\theta,\phi$)-independent matrix $U_R$:
\begin{equation} \label{eq:sph_to_mon}
	U_{\text{mon}}(\theta,\phi)=U_{\text{sph}}(\mu=\frac{\pi}{2},\theta,\phi).U_R,
\end{equation}
A key question is how we can test if $U_{\text{mon}}$ is correct for monopole field configurations. As derived in Eq.(\ref{eq::gauge_trans_Gauge}), the monopole gauge field configuration in KL convention read
\begin{equation}
\begin{aligned} \label{eq:Klinkhamer_gauge_f3_equal_f_monopole}
	A_{i}^aT^adx^i = \frac{1-f(\xi)}{g}iU_{\text{mon}}^{-1} (\partial_i U_{\text{mon}})dx^i,
\end{aligned}
\end{equation}
Substituting the expression \eqref{eq:sph_to_mon} into \eqref{eq:Klinkhamer_gauge_f3_equal_f_monopole}, the monopole gauge field becomes
\begin{equation} \label{eq:monopole_gauge_from_rotation_UR}
\begin{aligned}
A_{i}^aT^adx^i&= \frac{1-f(\xi)}{g}U_R^{-1}.\left[iU_{\text{sph}}^{-1} (\partial_i U_{\text{sph}})dx^i\right].U_R\\
&= \frac{1-f(\xi)}{g}U_R^{-1}\left[\sum_{i=1}^3F_aT^a \right]U_R\ \ \ .   
\end{aligned}
\end{equation}
Note that we do not introduce a additional matrix $U_L$ multiplying $U_{\text{sph}}(\mu=\frac{\pi}{2},\theta,\phi)$ on the left in Eq.~(\ref{eq:sph_to_mon}), since $U_L^{-1}.U_L=I$ always occurs in Eq.~(\ref{eq:Klinkhamer_gauge_f3_equal_f_monopole}). We now take the specific form for $U_R$ with\footnote{The $U_R$ matrix actually transforms the SU(2) generators into their complex conjugate representation, with $T^{a*}=-U_R^{-1}T^a U_R$, given in the appendix of Ref. \cite{Chao:2018xwz}.}:
\begin{equation} \label{eq:UR matrix}
	U_R=\left(\begin{array}{ccccc}
0 & 0 & \cdots & 0 & 1 \\
0 & 0 & \cdots & -1 & 0 \\
\cdots & \cdots & \cdots & \cdots & \cdots \\
0 & (-1)^{2 J-1} & \cdots & 0 & 0 \\
(-1)^{2 J} & 0 & \ldots & 0 & 0
\end{array}\right),
\end{equation}
where $J$ is the dimension of the scalar multiplet representation ({\it i.e.}, $J=1$ for the $Y=0$ complex triplet). Using this choice for $U_R$, we find that Eq.~(\ref{eq:monopole_gauge_from_rotation_UR}) leads to the gauge field configuration:
\begin{equation}
	A_i^a=\epsilon^{iab}x^b F\ \ \ ,
\end{equation}
where we have explicitly obtained this result for $J=[1,3/2,2,5/2,3]$; where
\begin{equation}
	F=\frac{2(-1+f(\xi))}{gr^2}\ \ \ ;
\end{equation}
and where this form is exactly the 't Hooft-Polyakov monopole gauge field configuration. 

To conclude this subsection, we discuss the field solution when both the SM Higgs and the multiplet obtain vevs. We parameterize the Higgs field, $H$, and the multiplet field, $\Phi$, as:
\begin{align} \label{eq:Higgs_Phi_locked_together_in_vacuum}
H=\frac{v}{\sqrt{2}}h(\xi).\begin{pmatrix}
    0 \\
    1
\end{pmatrix}, \quad \Phi=\frac{v_\phi}{\sqrt{2}}\phi(\xi).\begin{pmatrix}
    0 \\
    \cdots \\
    1 \\
    \cdots \\
    0
\end{pmatrix}.
\end{align}
where $v$ and $v_\phi$ represent the Higgs and multiplet vevs, and $h(\xi)$ and $\phi(\xi)$ are the Higgs and multiplet radial profile functions, respectively. We impose model parameter constraints to ensure that the configuration in Eq.~(\ref{eq:Higgs_Phi_locked_together_in_vacuum}) minimizes the potential. In this way, $\Phi$ is locked to the Higgs doublet at infinity, and also at finite radius through the radial profile functions (which can differ but must share the same boundary conditions). Therefore, the topology of the vacuum manifold remains $S^3$, ensuring that the sphaleron solution exists regardless of the hypercharge of $\Phi$.

\subsection{Mechanisms for sphaleron and monopole induced baryon number violation}

In this section, we review the roles of sphalerons and monopoles in the process of baryon number violation (BNV) and the corresponding expressions for the BNV rate. The fundamental principle underlying both sphaleron- and monopole-induced BNV is the Adler$-$Bell$-$Jackiw anomaly \cite{Adler:1969gk,Bell:1969ts}. For a theory  that conserves the baryon minus lepton number (like the electroweak theory), and for each fermion generation, the observable variation of baryonic and leptonic currents is given by
\begin{equation}
\label{eq:anomaly}
	\partial_\mu j^{\mu}_B=\partial_\mu j^{\mu}_L =\frac{g^2}{32\pi^2} F^{\mu \nu a} \widetilde{F}_{\mu \nu}^a,
\end{equation}
where $g$ is the SU(2)$_L$ coupling constant and $\widetilde{F}_{\mu \nu}^a=\frac{1}{2}\epsilon_{\mu \nu \alpha \beta}F^{\alpha \beta a}$. The right hand side of Eq.(\ref{eq:anomaly}) can be written as a total derivative
\begin{align} \label{eq:Kmu_definition}
    \frac{g^2}{32\pi^2} F^{\mu \nu a} \widetilde{F}_{\mu \nu}^a = \partial_\mu K^\mu,
\end{align}
where $K^\mu$ is defined as
\begin{align} \label{eq:K_mu_definition}
    K^{\mu}\equiv \frac{g^{2}}{32\pi^{2}}\epsilon^{\mu \nu \rho \sigma} (F_{\nu \rho}^{a}A_{\sigma}^{a}-\frac{g}{3}f^{abc}A_{\nu}^{a}A_{\rho}^{b}A_{\sigma}^{c}) ,
\end{align}
with $f^{abc}$ being the non-abelian gauge group structure constant and $F_{\nu\rho}^a$ being the gauge field strength tensor
\begin{equation}
	F_{\nu \rho}^{a}=\partial_{\nu}A_{\rho}^{a}-\partial_{\rho}A_{\nu}^{a}+gf^{ade}A_{\nu}^{d}A_{\rho}^{e} \ \ .
\end{equation}
We further define a function $\mathcal{P}(t_a,t_b)$ of general initial and final time boundaries, denoted by $t_a$ and $t_b$
\begin{equation}
\begin{aligned} \label{eq:Ptatb_definition}
	\mathcal{P}(t_a,t_b)&\equiv\int_{t_a}^{t_b} dt \int_{-\infty}^{+\infty} d^3 x\ \partial_\mu K^\mu \\
	&=\int d^{3} x K^{0}\bigg\vert^{ t=t_b}_{ t=t_a}+\int^{t_b}_{t_a}dt \int_{S} \vec{K}\cdot \vec{dS}\ , 
\end{aligned}
\end{equation} 
where the spatial part is integrated over all space, and where we applied the divergence theorem to $K^i$ in the second line.

The winding number $\omega$ can be defined using $\mathcal{P}(t_a,t_b)$: 
\begin{equation} \label{eq:winding_number_definition}
	\omega \equiv \mathcal{P}(t_a=-\infty,t_b=+\infty).
\end{equation}
which equals an integer for a dynamic process starting and ending at topologically distinct vacua. 

\paragraph{The sphaleron induced baryon number violation} 
We start with definition of the sphaleron baryonic charge. 
According to Eq.(\ref{eq:anomaly},\ref{eq:Kmu_definition},\ref{eq:Ptatb_definition}), the $\mathcal{P}(t_a,t_b)$ has the following relation with baryonic current $j_B^\mu$,
\begin{equation}
    \mathcal{P}(t_a,t_b) = \int d^3 x\ j_B^0\bigg\vert^{t_b}_{t_a}+\int_{t_a}^{t_b} dt \int_{-\infty}^{+\infty} d^3 x\ \nabla\cdot \vec{j}_B,
\end{equation}
The sphaleron baryonic charge $Q_B$ is then defined as
\begin{equation} \label{eq:sphaleron_baryonic_charge_definition}
\begin{aligned}
    Q_B &\equiv \int d^3 x\ j_B^0\bigg\vert^{t_0}_{-\infty} \\
    &=-\int_{-\infty}^{t_0} dt \int d^3 x\ \nabla\cdot \vec{j}_B + \int_{-\infty}^{t_0} dt \int d^3 x\ \partial_\mu K^\mu,
\end{aligned}
\end{equation}
where $t_0$ represents the sphaleron point corresponding to the top of the energy barrier in configuration space, connecting topologically distinct vacua. Assuming as in \cite{Klinkhamer:1984di} that ${\vec j}_B$ exhibits only standard (topologically trivial) properties, 
we can neglect the $\nabla\cdot \vec{j}_B$ term in the baryonic charge definition. Henceforth, in evaluating $Q_B$ (Eq.~(\ref{eq:sphaleron_baryonic_charge_definition})), we shall restrict our attention to the contribution arising from $\partial_\mu K^\mu$, which is topologically nontrivial. We will show that the $Q_B$ takes half-integer values.

In the sphaleron configuration, the non-contractible loop parameter $\mu$ is a dynamic variable, which takes different values along the time variable entering $\mathcal{P}(t_a,t_b)$. Regarding $\mu$ as a monotonically increasing function of $t$ \cite{Tye:2015tva,Tye:2016pxi}, we can replace the time boundaries in $\mathcal{P}(t_a,t_b)$ with the corresponding values of $\mu$: $\mathcal{P}(\mu(t_a),\mu(t_b))$. Thus, the variation of $\mu$ in sphaleron field configuration leads into variation of winding number and sphaleron baryonic charge.

Now we will demonstrate the sphaleron half-integer baryonic charge property. According to Eq.(\ref{eq:sphaleron_baryonic_charge_definition},\ref{eq:Ptatb_definition}), the sphaleron bartonic charge $Q_B$ reads
\begin{equation}
\begin{aligned} \label{eq:Q_B sphaleron definition new}
Q_{B}&=\int d^{3} x K^{0}\bigg\vert^{t=t_{0}}_{t=-\infty}+\int^{t_{0}}_{-\infty}dt \int d^3 x\, \nabla\cdot \vec{K}.
\end{aligned}
\end{equation}
At $t=-\infty$, the gauge fields are in vacuum states, while at $t=t_0$ their configuration corresponds to being at the top of the field energy barrier separating two adjacent topologically distinct vacua. We will compute both the first $K^0$ term and the second surface term in Eq.~\eqref{eq:Q_B sphaleron definition new} using the KL configuration, where the gauge field falls off as $1/r$. 
In the KL configuration, we choose:
\begin{equation} \label{eq:mu_relation_with_t_vacuum_saddle}
    t=-\infty,\ \mu=0;\ \  t=t_0,\ \mu=\frac{\pi}{2};\ \  t=\infty, \ \mu=\pi,
\end{equation}
where the same relation of  $\mu$ and $t$ also applies for the MK configuration. Note that the definition of $Q_B$ is only related with the gauge field, so phases I and III of the KL configuration, wherein only the scalar field varies, are not relevant (see 
Section \ref{sec:Sphaleron energy, monopole mass and equation of motions} below for a discussion of the KL configuration).

Now we label the sphaleron baryonic charge in the KL configuration as $Q_B(\text{KL})$. Under this convention, the gauge field is given by
\begin{align} \label{eq:KL_gauge_confg_in_QB}
&A_i^a({\rm KL}) T^a dx^i = \frac{1}{g}(1-f(\xi))\left[F_1 T^1+F_2 T^2\right]  \nonumber \\
    & \ \  \ \ \ \ \ \ \  \qquad \qquad +\frac{1}{g}(1-f_3(\xi))F_3 T^3 ,
\end{align}
where $f(\xi)$ and $f_3(\xi)$ are the radial profile functions of the SU(2) gauge fields; and the one-form $F_a$ are defined in Eq.~(\ref{eq:one_form_definition}). Substituting this gauge field configuration into $K^\mu$ (Eq.~(\ref{eq:K_mu_definition})) yields $K^0=0$. 
Consequently, the  $\nabla\cdot \vec{K}$ term in $Q_B$ alone generates the entire baryonic charge:
\begin{equation} \label{eq:Q_B_KL_convention}
\begin{aligned}
&Q_B(\text{KL})=\int^{t_{0}}_{-\infty}dt \int d^3x\, \nabla\cdot \vec{K}\\
&\quad =-\int _{-\infty}^{t_{0}}dt \frac{4}{3\pi}\left[(-1+f(\xi))^{2}+\frac{1}{2}(-1+f_{3}(\xi))^{2}   \right] \\
&\qquad \qquad \times \sin^{2}(\mu(t))\frac{d\mu(t)}{dt}\bigg\vert^{r=\infty}_{r=0}  \ \ ,
\end{aligned}
\end{equation}
Here, we have (i) expanded $\nabla\cdot \vec{K}$ in spherical coordinates, (ii) substituted the field configuration (Eq.~(\ref{eq:KL_gauge_confg_in_QB})) into $K^i$. By imposing the boundary conditions $f(0)=f_3(0)=0$ and $f(\infty)=f_3(\infty)=1$ (see Eq.(\ref{eq:sphaleron_boundary_conditions})), one obtains
\begin{equation}
\begin{aligned}
    Q_B(\text{KL})& =\frac{1}{\pi}(\mu-\frac{1}{2}\sin(2\mu))\big\vert^{\mu=\frac{\pi}{2}}_{\mu=0}=\frac{1}{2}\ \ \ .
\end{aligned}
\end{equation}
Ref.~\cite{Klinkhamer:1984di} argues that $Q_B$ is gauge-invariant, so that its value should remain unchanged even in other sphaleron configurations, such as the MK 
configuration, that are related to the KL configuration through a gauge transformation. Through explicit computation, we find that this argument holds up to an overall sign, and that evidently only the absolute value of $Q_B$ stays invariant, {\it i.e.} $|Q_B|=1/2$. 
For example, in the MK configuration (see Eq.~(\ref{eq:sphaleron_gauge_confg_manton})), 
The value of $Q_B$ reads (again, $K^0=0$)
\begin{equation} \label{eq:Q_B_MK_convention}
\begin{aligned}
&Q_B(\text{MK})=\int^{t_{0}}_{-\infty}dt \int d^3x\, \nabla\cdot \vec{K}\\
&\quad =-\int _{-\infty}^{t_{0}}dt \frac{2}{\pi} f^2(\xi) \times \sin^{2}(\mu(t))\frac{d\mu(t)}{dt}\bigg\vert^{r=\infty}_{r=0},  
\end{aligned}
\end{equation}
where we follow the same computational strategy as for $Q_B({\rm KL})$, but now impose the MK constraint $f_3(\xi)=f(\xi)$. Finally, applying the boundary conditions $f(0)=0$ and $f(\infty)=1$, yields
\begin{equation}
\begin{aligned}
Q_B(\text{MK}) =-\frac{1}{\pi}(\mu-\frac{1}{2}\sin(2\mu))\big\vert^{\mu=\frac{\pi}{2}}_{\mu=0}=-\frac{1}{2}\ \ \ .
\end{aligned}
\end{equation}
  Compare the second line of Eq.~(\ref{eq:Q_B_KL_convention}) and Eq.~(\ref{eq:Q_B_MK_convention}), we see that the expression of $Q_B$ is different in these two configurations\footnote{For KL configuration with $f_3(\xi)=f(\xi)$, we will obtain a profile factor of $\frac{2}{\pi}(-1+f(\xi))^2$ in the second line of $Q_B(\text{KL})$, which is different with that in MK case, the $\frac{2}{\pi}f^2(\xi)$. }. Moreover, its value depends explicitly on the choice of field boundary conditions. A complete analysis of the boundary conditions is carried out in later Sec.~\ref{sec:Sphaleron energy, monopole mass and equation of motions}, where we show that our choice of boundary conditions used here is unique in order to obtain a finite sphaleron energy. Such a sign difference may or may not affect the BNV process, and we deter such implication into future studies.

We also note that, as demonstrated in Ref.~\cite{Klinkhamer:1984di}, an additional gauge transformation (\yd{$U_c$}) can be applied to make the gauge field falls off faster than $1/r$ so that the surface term in Eq.~\eqref{eq:Q_B sphaleron definition new} vanishes,
\begin{equation} \label{eq:Uc_definition_MAIN_text}
U_c=\exp((-1)^n i\Theta(r) \vec{T}\cdot \hat{\vec{x}}),
\end{equation}
where $n=1$ or $2$ regarding on the convention of sign; where $\Theta(r)$ runs rapidly from $0$ to $\pi$ when $r$ runs from $0$ to $\infty$; 
where $\vec{T}$ are the SU(2) generators, and $\hat{\vec{x}}=\vec{x}/r$. In such a gauge, the non-vanishing $Q_B$ arises from the first term in Eq.~\eqref{eq:Q_B sphaleron definition new}. In this scenario (see Appendix \ref{app:gauge_choice_Aia_falls_faster_1over_r}), the MK configuration produces $Q_B=+\frac12$ for $n=1$ and $Q_B=-\frac12$ for $n=2$.    

To end this subsection, we now recall the rate for BNV violation associated with transitions between adjacent vacua that pass through the configuration with $Q_B=1/2$. Following Ref.~\cite{Patel:2011th} (and references therein) we write
\begin{equation} \label{eq:sphaleron_BNV_expression}
\frac{\partial \bar{n}_B}{\partial t}+3H\bar{n}_B= - k(t) \bar{n}_B,
\end{equation}
where $\bar{n}_B$ is the net baryon number density, $\bar{n}_B\equiv n_B-n_{\bar{B}}$, $H$ is the Hubble constant, and
\begin{equation}
k(t) = \frac{13 n_f}{2} \frac{\Gamma_\mathrm{sph}}{VT^3} ,
\end{equation}
with $n_f$ being the number of fermion generations and $V$ being the space volume; where 
\begin{align} \label{eq:Gamma_sph_definition}
\Gamma_{\text{sph}}=A_{\text{sph}}e^{-E_{\text{sph}}/T}.
\end{align}
Here,  $E_{\text{sph}}/T$ is the leading order action evaluated at the sphaleron point; $A_{\text{sph}}$ collects the quantum fluctuations and represents the higher-order corrections to the sphaleron rate. 
We will elaborate the discussion of $E_\text{sph}$ after the corresponding discussion of the monopole-induced BNV.

\paragraph{Monopole induced baryon number violation} In contrast to the sphaleron, the monopole is a topologically stable configuration. In the monopole background, the following scattering process can occur\footnote{To simplify the presentation, we have omitted the chirality and potential flavor quantum numbers of quarks and leptons.},
\begin{equation}
    u+u\rightarrow \bar{d}+e^+\ \ ,
\end{equation}
where $u$, $\bar{d}$, and $e^+$ denote the up quark, anti-down quark, and position, respectively. This process changes the baryon plus lepton number by $\Delta (B+L)=-2$. As demonstrated by Rubakov and Callan \cite{Rubakov:1981rg,Rubakov:1982fp,Callan:1982ac}, it is the non-static fluctuation fields around the static monopole solution give that rise to such $B+L$ violation scattering. We write these quantum fluctuations as $a_\mu(r,t)$, with:
\begin{equation}
\begin{aligned}
A_0&=T^a \hat{x}^a a_0(r,t)/i,\\
A_i&=T^a \hat{x}^a  \hat{x}_i a_1(r,t)/i+A_i^{\text{cl}}\ \ \  ,\\
\Phi &= \Phi^{\text{cl}}\ \ \ ,
\end{aligned}
\end{equation}
where $T^a$ are the SU(2) generators, $ \hat{\vec{x}}=\vec{x}/r$, $A_i^{\text{cl}}$ and $\Phi^{\text{cl}}$ are the classical background monopole solutions following the notations in Refs.~\cite{Rubakov:1982fp}. The fluctuation fields $a_0(r,t)$ and $a_1(r,t)$ obey:
\begin{align}
    a_0(r,\pm \infty)=a_1(r,\pm \infty)=0\ \ \ .
\end{align}
Importantly, there exist certain fluctuation configurations obeying these boundary conditions that (1) yield the integer winding number; and (2)
lead the action functional to be arbitrary small. The second property implies that the fermion number non-conserving matrix element is not suppressed by a factor of $\exp(-\text{const}/\text{coupling}^2)$ \cite{Rubakov:1982fp}. 

The monopole induced BNV comes from the scattering process of the $s-$wave fermions and monopole, with corresponding cross-section denoted as $\sigma_{\Delta B\neq 0}$. 
In general, one would expect this cross section to depend on the monopole size, the scattering kinematics ({\it e.g.}, center of mass scattering energy), and possibly a non-perturbative exponential function of the gauge coupling, $g$.
Refs.~\cite{Rubakov:1981rg,Rubakov:1982fp,Callan:1982ac} show that $\sigma_{\Delta B\neq 0}$ is not suppressed either by the geometric size of the monopole or by a non-perturbative factor exp($-f(v)$/$g^2$), with $f(v)$ being some constant function of the the multiplet vev, $v$. Denoting the remaining relevant energy scale as the \lq\lq fermion energy\rq\rq, $E_f$, one may thus write \cite{Ellis:1982bz}
\begin{align}
\sigma_{\Delta B\neq 0} \simeq v^{-1}\frac{c}{E_f^2}.
\end{align}
where $v$ is the monopole-fermion relative velocity and $c$ is an appropriate constant. 
This $\sigma_{\Delta B\neq 0}$ plays a similar BNV role as that of $\Gamma_\text{sph}$ in the sphaleron case as encoded in the right hand side of Eq.~(\ref{eq:sphaleron_BNV_expression}), although they have distinct features that we now delineate. 

For a Boltzmann equation description of monopole-catalyzed BNV, analogous to Eq.~(\ref{eq:sphaleron_BNV_expression}),
one requires in addition to the cross section the monopole and fermion number densities in the early universe plasma. To arrive at the former, we observe that 
during the period of EWSB, monopoles can be created through the bubble collision mechanism \cite{Preskill:1984gd} (if EWSB occurs via a first order phase transition) or the monopole thermal pair production mechanism \cite{Patel:2012pi}. Henceforth, we assume a first order EWPT. It has been estimated that the thermal monopole anti-monopole pair production rate exceeds the bubble collision rate over a wide range of parameters \cite{Patel:2012pi}. The thermal monopole anti-monopole pair production mechanism leads to an equilibrium monopole density of
\begin{equation} \label{eq:monopole_thermal_pair_production_thermal_equilibrium}
	\frac{n_M}{T^3}\bigg\vert_{\text{eq.}}=\left(\frac{m_M(T)/T_{\text{nuc}}}{2\pi} \right)^{3/2}e^{-m_M(T)/T},
\end{equation}
where $n_M$ is the monopole density, $m_M(T)$ is the monopole mass, and $T_{\text{nuc}}$ is the nucleation temperature. 

The resulting Boltzmann equation for monopole-catalyzed BNV is given by
\begin{align} \label{eq:monopole_boltzmann_eq}
    \frac{\partial \bar{n}_B}{\partial t}+3H\bar{n}_B = -\langle\sigma_{\Delta B\neq 0} v\rangle g_f \bar{n}_f n_M \  \ ,
\end{align}
which is derived in Appendix.~\ref{app:Boltzmann_eq_of_monopole_fermion}; where $\langle \ldots \rangle$ denotes the thermal average of the enclosed quantity, $g_f$ is the number of fermion degrees of freedom, $n_f$ is the net fermion number density, $\bar{n}_f\equiv n_f - n_{\bar{f}}$, and $n_M$ is the monopole density. In this study, we assume the monopole density is in thermal equilibrium, given by Eq.(\ref{eq:monopole_thermal_pair_production_thermal_equilibrium}), where an exponential suppression factor $e^{-m_M/T}$ appears.

In summary, for our current leading-order BNV rate computation, we focus only on the computation of the exponential factor, $e^{-E/T}$, where $E$ refers to either sphaleron energy or monopole mass. We stress that the origin of $e^{-E/T}$ is different in these two cases. For the sphaleron case it comes from the sphaleron transition rate between adjacent vacua, while in the monopole case it arises from the equilibrium monopole density. We defer an exact computation of the monopole density as well as next-to-leading-order contribution to the sphaleron rate or monopole-fermion BNV scattering cross-section to future work. 

\subsection{Sphaleron energy, monopole mass and equations of motion} \label{sec:Sphaleron energy, monopole mass and equation of motions}

Our computation in this subsection is based on the KL sphaleron configuration, from which one may also derive the monopole configuration
according to Eq.(\ref{eq:sph_to_mon}). To enhance the  accuracy of the sphaleron energy computation, we include the contribution from the U(1) gauge field\footnote{An additional motivation for including the $U(1)$ gauge field is its ability to impart an elliptic deformation to the otherwise spherical sphaleron explosion, thereby generating gravitational radiation \cite{Kharzeev:2019rsy}.}. For the monopole mass calculation, which pertains to a $Y=0$ multiplet extension, the U(1) gauge field 
decouples and does not contribute at leading order. Nevertheless, we still include the U(1) field in the monopole configuration at the beginning as a more general setup and show explicitly how it decouples when $Y=0$. 

\paragraph{Sphaleron energy.} The non-contractible loop (NCL) in the KL configuration commences and terminates at topologically distinct vacua and comprises three phases \cite{Klinkhamer:1990fi}
\begin{itemize} \label{item:KL_confg_three_phases}
    \item I, $\mu \in [-\pi/2,0]$: builds up the scalar field configuration;
    \item II, $\mu \in [0,\pi]$: builds up and then destroys the gauge field configuration;
    \item III, $\mu \in [\pi, 3\pi/2]$: destroys the scalar field configuration.
\end{itemize}
where $\mu=-\pi/2,\ 3\pi/2$ represent the vacuum state, while $\mu=\pi/2$ denotes the sphaleron state. The field ansatzes vary in different phases. In phase I and III, the profile functions are given by:\footnote{In general, the gauge fields need not vanish in phases I and III except at $\mu=-\tfrac{\pi}{2}$ and $\mu=\tfrac{3\pi}{2}$. Here, we impose that the gauge fields are nonzero only in phase II \cite{Klinkhamer:1990fi}. Although this restriction is not fully general, it simplifies the topological discussion in the preceding subsections and the computation of $Q_B$ in Eq.~(\ref{eq:Q_B_KL_convention}), since the gauge fields then exist only for $\mu\in[0,\pi]$ under both the KL and MK conventions.}
\begin{align} \label{eq::Klinkharmer_gound_gauge_new}
    &A_i^aT^a=a_i=0, \\
\label{eq::Klinkharmer_gound_scalar_new}
    &\Phi = \frac{v_\phi\left(\sin ^2 \mu+\phi(\xi) \cos ^2 \mu\right)}{\sqrt{2}}\left(\begin{array}{c}
     0 \\
     \cdots \\
     1 \\
     \cdots \\
     0
\end{array} \right)\ \ \ ,
\end{align}
where $a_i$ denotes the $U(1)$ gauge field. In phase II, the field configurations read:
\begin{align}
    \label{eq:multiplet_sphaleron_confg_KL_scalar}
        &\Phi=\frac{v_\phi}{\sqrt{2}}\phi(\xi)\left(\begin{array}{c}
             0 \\
            \cdots \\
            1 \\
            \cdots \\
            0
        \end{array} \right),\\
        \label{eq:multiplet_sphaleron_confg_KL_gauge}
        &A_i^a T^a dx^i = \frac{1}{g}(1-f(\xi))\left[F_1 T^1+F_2 T^2\right] \\
        & \qquad \qquad+\frac{1}{g}(1-f_3(\xi))F_3 T^3, \nonumber \\
        \label{eq:multiplet_sphaleron_confg_KL_U1_field}
        & a_i dx^i = \frac{1}{g^\prime}(1-f_0(\xi))F_3\ \ \ .
\end{align}
where $g^\prime$ denotes the $U(1)$ gauge coupling constant; $f$ and $f_3$ are the SU(2) gauge field radial profile functions; $f_0$ is the U(1) gauge field radial profile function; $F_1$,$F_2$ and $F_3$ are functions of $\theta$ and $\phi$ defined in Eq.~(\ref{eq:one_form_Fa_definition}).

In the following analysis, we construct the sphaleron energy using a single scalar multiplet extension to the SM, while referring to 
$V(H,\Phi)$ as a general object to demonstrate the main ideas. We define that $V(H,\Phi)$ vanishes in the limit $\xi\rightarrow\infty$. We will present the explicit interaction terms in the next section. The field energy can be written as
\begin{equation}
\begin{aligned} \label{eq::formal_sphaleron_formula}
        E&=\frac{4\pi \Omega}{g}\int d\xi \bigg[ \frac{1}{4}F^a_{ij}F^a_{ij}+\frac{1}{4}f^a_{ij}f^a_{ij}+ (D_iH)^{\dagger}(D_iH)  \\
        & \quad \quad\quad\quad\quad\quad  + (D_i\Phi)^{\dagger}(D_i\Phi) + V(H,\Phi) \bigg],
\end{aligned}
\end{equation}
where $\Omega$ is the zero temperature vev for the SM Higgs field. The general expressions of Yang-Mills term, $U(1)$ term and covariant derivative term read
\begin{widetext}
\begin{equation} \label{eq:sphaleron_energy_density_terms}
\begin{aligned}
\frac{1}{4}F_{ij}^{a}F_{ij}^{a}(\xi,\mu) &= \sin ^{2} \mu\left(\frac{8}{3} f^{\prime 2}+\frac{4}{3} f_{3}^{\prime 2}\right)+\frac{8}{\xi^{2}} \sin ^{4} \mu\left\{\frac{2}{3} f_{3}^{2}(1-f)^{2}+\frac{1}{3}\left\{f(2-f)-f_{3}\right\}^{2}\right\} ,\\
\frac{1}{4}f_{ij}f_{ij}(\xi,\mu)&=\frac{4}{3}\left(\frac{g}{g^{\prime}}\right)^{2}\left\{f_{0}^{\prime 2} \sin ^{2} \mu +\frac{2}{\xi^{2}} \sin ^{4} \mu\left(1-f_{0}\right)^{2}\right\} ,\\
(D_{i}\Phi)^{\dagger}(D_{i}\Phi)(\xi,\mu)&=\frac{v^{2}_\phi}{\Omega^{2}}\left\{\frac{1}{2} \xi^{2} \phi^{\prime 2}+\frac{4}{3} \phi^{2} \sin ^{2} \mu \left\{\left(J(J+1)-J_{3}^{2}\right)(1-f)^{2}+J_{3}^{2}\left(f_{0}-f_{3}\right)^{2}\right\}\right\} \ \ \ .
\end{aligned}
\end{equation}
\end{widetext}
Note that above expressions hold for phase II of KL configurations, where $\mu\in [0,\pi]$. The Higgs covariant derivative term $(D_iH)^\dagger (D_i H)$ in Eq.~(\ref{eq::formal_sphaleron_formula}) can be obtained from the expression of $(D_i\Phi)^\dagger (D_i \Phi)$ in Eq.~(\ref{eq:sphaleron_energy_density_terms}) with the choice of $J=1/2$ and $J_3=1/2$ and the  field notation $\phi\rightarrow h$. 
Thus, the sphaleron energy can be expressed as
\begin{equation} \label{eq::sphaleron_energy}
    \begin{aligned}
        E_{\text{sph}}&= E(\mu=\frac{\pi}{2}) \\
        & = \frac{4\pi \Omega}{g}\int_0^\infty d\xi \left[ \frac{1}{4}F_{ij}^aF_{ij}^a(\xi,\frac{\pi}{2}) + \frac{1}{4}f_{ij}f_{ij}(\xi,\frac{\pi}{2}) \right. \\
        & \left.\quad + (D_iH)^{\dagger}(D_iH) (\xi,\frac{\pi}{2})  + (D_i\Phi)^{\dagger}(D_i\Phi)(\xi,\frac{\pi}{2}) \right. \\ &\left. \quad  + V(H,\Phi)(\xi,\frac{\pi}{2})\right].
    \end{aligned}
\end{equation}

\paragraph{Monopole mass} Recall that for the monopole, its field configuration is stable, and there is no need for an NCL. According to Eq.(\ref{eq:monopole_gauge_from_rotation_UR}), the monopole configuration can be constructed from the sphaleron configuration in phase II when $\mu=\pi/2$, {\em viz}
\begin{equation} \label{eq::Multiplet_monopole_confg_KL}
    \begin{aligned}
        &\Phi=\frac{v_\phi}{\sqrt{2}}\phi(\xi)\left(\begin{array}{c}
             0 \\
            \cdots \\
            1 \\
            \cdots \\
            0
        \end{array} \right),\\
        &A_i^a T^a dx^i = \frac{1}{g}(1-f(\xi))U_R^{-1}.\left[F_1 T^{1}+F_2 T^{2}\right].U_R\big\vert_{\mu=\frac{\pi}{2}} \\
        & \qquad \qquad-\frac{1}{g}(1-f_3(\xi))U_R^{-1}.F_3 T^{3}.U_R\big\vert_{\mu=\frac{\pi}{2}}, \\
        & a_i dx^i = \frac{1}{g^\prime}(1-f_0(\xi))F_3\big\vert_{\mu=\frac{\pi}{2}}\ \ \ .
    \end{aligned}
\end{equation}
where $U_R$ is a constant matrix defined in Eq.~(\ref{eq:UR matrix}); the $\mu=\pi/2$ condition is applied to the 1-form $F_a$. 

Now we discuss the definition of monopole mass. When $\mu=\frac{\pi}{2}$, we find that the forms of the monopole field gauge configurations lead to Yang-Mills and $U(1)$ terms that are identical to those of the sphaleron case since the $U_R$ rotation matrix has no effect on these two terms. On the other hand, for general values of $J$ and $J_3$, the presence of the $U_R$ matrix would lead to a different scalar field kinetic term compared to the sphaleron case. However, the difference comes only from the coefficient of $J_3$. For a $Y=0$ multiplet, the neutral component that appears in the monopole solution has $J_3=0$. For this case, one may also obtain the scalar kinetic term from the corresponding sphaleron case by setting $J_3=0$ and $Y=0$.
Thus, the monopole mass can be defined as:
\begin{equation} \label{eq::monopole_mass_explicit_expression}
\begin{aligned}
m_M&\equiv \frac{4\pi \Omega}{g}\int_0^\infty d\xi
    \left[ \frac{1}{4}F_{ij}^aF_{ij}^a(\xi,\frac{\pi}{2}) + \frac{1}{4}f_{ij}f_{ij}(\xi,\frac{\pi}{2}) \right. \\
        & \left.\quad   + (D_i\Phi)^{\dagger}(D_i\Phi)(\xi,\frac{\pi}{2}) + V(H,\Phi)(\xi,\frac{\pi}{2}) \right].
\end{aligned} 
\end{equation}

In what follows, we present the  equations of motion and corresponding boundary conditions as needed to obtain the sphaleron energy and monopole mass. Subsequently in Sec.~\ref{sec:sphaleron_energy}, where we numerically solve the monopole field equations of motion and present the result in Fig.\ref{fig:One_and_Two_EWPT_profile_function}(b), we see that $f_0(\xi)=1$ for all range of $\xi$, so the U(1) field is totally decoupled as expected for the monopole case.

\paragraph{Equation of motion.}
For a general setup, we consider the computation of sphaleron equations of motion when both the SM Higgs and an additional multiplet obtain vevs, which corresponds to sphaleron energy at point $Z$ in Fig.~\ref{fig:different_steps_EWPT}. The sphaleron or monopole equations of motion at points $X$ and $W$ can be obtained through simplifications of this general setup.

The equations of motion (EOM) are differential equations governing the radial profile functions of the field ansatzes: $f$ and $f_3$ for the SU(2) gauge field; $f_0$ for the U(1) gauge field; $h$ for the SM Higgs field; and $\phi$ for the additional multiplet field. Similar to the case in Ahriche et al.'s work \cite{Ahriche:2014jna}, the EOMs reads 
\begin{widetext}
\begin{equation} \label{eq::EOMs}
\begin{aligned}
f^{\prime \prime}+\frac{2}{\xi^{2}}(1-f)\left[f(f-2)+f_{3}\left(1+f_{3}\right)\right]+(1-f)(\frac{v^{2}h^{2}}{4\Omega^{2}}+\alpha \phi^{2})=0, \\
f_{3}^{\prime \prime}-\frac{2}{\xi^{2}}\left[3 f_{3}+f(f-2)\left(1+2 f_{3}\right)\right]+(\frac{v^{2}}{4\Omega^{2}}h^{2}+\beta \phi^{2})(f_{0}-f_{3})=0, \\
f_{0}^{\prime \prime}+\frac{2}{\xi^{2}}\left(1-f_{0}\right)-\frac{g^{\prime 2}}{g^{2}}(\frac{v^{2}}{4\Omega^{2}}h^{2}+\beta \phi^{2})(f_{0}-f_{3})=0, \\
h^{\prime \prime}+\frac{2}{\xi}h^{\prime}-\frac{2}{3\xi^{2}}h[2(1-f)^{2}+(f_{0}-f_{3})^{2}]-\frac{1}{g^{2} v^{2} \Omega^{2}}\frac{\partial V[h,\phi]}{\partial h}=0,\\
\phi^{\prime \prime}+\frac{2}{\xi}\phi^{\prime}-\frac{8\Omega^{2}\phi}{3 v_{\phi}^{2} \xi^{2}}[2\alpha(1-f)^{2}+\beta (f_{0}-f_{3})^{2}]-\frac{1}{g^{2} v_{\phi}^{2} \Omega^{2}}\frac{\partial V[h,\phi]}{\partial \phi}=0,
\end{aligned}
\end{equation}
\end{widetext}
where $f^{\prime}$ denotes $df/d\xi$ and $f^{\prime \prime}$ denotes $d^2 f/d\xi^2$. In the zero temperature computation, we set $v=\Omega=246.22$ GeV. However, at finite temperatures, $v$ is a function of the temperature; and $\alpha$ and $\beta$ are defined as
\begin{equation}
\alpha = \frac{[J(J+1)-J_{3}^{2}]v_{\phi}^{2}}{2 \Omega^{2}},\quad \beta = \frac{J_{3}^{2}v_{\phi}^{2}}{\Omega^{2}}.
\end{equation}
where $J$ denotes the multiplet representation dimension, and $J_3$ is the third component value. Since we put the multiplet's vev in its neutral component, $J_3$ equals to the opposite value of hypercharge $Y$.
The only as yet unspecified term in the EOMs in Eq.~\eqref{eq::EOMs} is the potential term $V[h,\phi]$, which depends on the specific BSM model and type of EWPT. 

Before concluding this section, we will clarify some subtleties regarding the EOM boundary conditions. The boundary condition for  fields at spatial infinity is clear: each field should approach its vacuum. On the other hand, at the spatial origin, some subtleties would appear, depending on the choice of co-ordinate system. Working with spherical-polar co-ordinates, the usual criteria for boundary conditions can be summarized as \cite{Manton:1983nd,Klinkhamer:1984di}
\begin{itemize}
    \item when $\xi\rightarrow 0$, the fields are singularity-free\ \ ,
    \item when $\xi\rightarrow \infty$, the fields asymptotically approach to the \lq\lq vacuum state\rq\rq. The \lq\lq vacuum state\rq\rq refers to the gauge and scalar field configurations making the Yang-Mills term and potential term vanish, and any derivative terms in the Hamiltonian need to vanish asymptotically for a finite energy configuration. 
\end{itemize}
Above two criteria work well in the MK sphaleron configuration. One might wonder whether they still hold in the KL configuration, as given in Eq.~(\ref{eq:multiplet_sphaleron_confg_KL_gauge}). In particular, the first singularity-free condition at the origin requires $f(\xi)=f_3(\xi)=1$ as $\xi\rightarrow0$. We will shortly show that this requirement conflicts with the sphaleron finite energy condition. Therefore, it is necessary to investigate the boundary conditions of the equations of motion in detail.
In this work, we will show that the first boundary condition is not necessary; the only essential condition comes from the finite energy requirement. 

Let us elaborate on the singularity issue. According to the MK configuration Eq.~\eqref{eq:sphaleron_gauge_confg_manton}, $U_{\text{sph}}$ is a function that contains the angular parameters $\theta,\phi$. When $r\rightarrow 0$, if the field does not vanish, the field would have some preferred angular direction at the origin, which can lead to a rotational singularity. In the following, we will demonstrate that such singularity is removable. As demonstrated in Eq.~(\ref{eq::gauge_trans_Gauge}), a unitary gauge transformation can connect following two field configurations under the zero weak mixing angle scenario
\begin{equation} \label{eq:unitary_transformation_slip_boundary_condition}
    -f(\xi) (\partial_i U)U^{-1} \overset{U}{\longleftrightarrow} [1-f(\xi)] U^{-1}\partial_i U,
\end{equation}
which means such a gauge transformation can interchange the boundary condition at the origin and spatial infinity. For example, following two sets of boundary conditions can be converted to each other by such a gauge transformation: 
\begin{itemize} 
    \item (a) $\xi\rightarrow 0$, $f(\xi)\rightarrow 0$; $\xi\rightarrow \infty$, $f(\xi)\rightarrow 1$; 
    \item (b) $\xi\rightarrow 0$, $f(\xi)\rightarrow 1$; $\xi\rightarrow \infty$, $f(\xi)\rightarrow 0$.
\end{itemize}

\begin{figure}[t]
\center
\includegraphics[width=7cm]{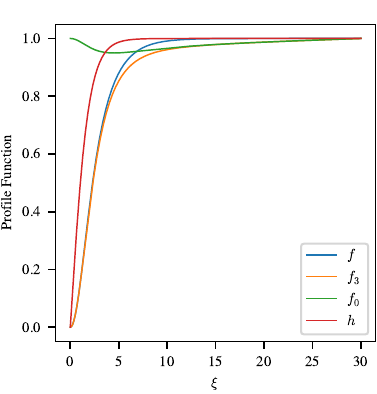}
\caption{The solutions of profile functions in the standard model. The set of equation of motions \eqref{eq::EOMs} are solved using the $\texttt{solve\_bvp}$ function from the open-source Python package $\texttt{scipy.integrate}$. 
The horizontal axis is a dimensionless radial parameter $\xi = g \Omega r$, where $\Omega=246$ GeV and $g$ is the SU(2) gauge coupling constant. The corresponding sphaleron energy $B_\text{sph}=1.900506$, where the formal expression $E_{\text{sph}}=B_\text{sph}\cdot(4 \pi \Omega)/g$ is defined in Eq.~\eqref{eq::sphaleron_B_definition}.
}
\label{fig:SM_different_profile}
\end{figure}

Thus, when $\xi \rightarrow 0$, the free of singularity condition is not strict. Since we can always make such gauge transformation to remove the singularity. In fact, the only criterion for sphaleron or monopole field boundary condition is 
\begin{itemize}
    \item sphaleron/monopole has finite energy
\end{itemize}
In the following, we use the sphaleron example for illustration. The sphaleron energy is an integral over dimensionless radial parameter $\xi$, given in Eq.(\ref{eq::formal_sphaleron_formula}), with the energy density in different terms given in Eq.(\ref{eq:sphaleron_energy_density_terms}). To ensure a finite sphaleron energy, we require
\begin{itemize}
    \item (i) when $\xi \rightarrow 0$, all energy density components have finite values.
    \item (ii) when $\xi \rightarrow \infty$, all energy density components asymptotically approach zero, decaying more rapidly than $1/\xi$.
\end{itemize}
The criterion (i) requires the coefficient proportional to $1/\xi^2$ vanishes in Eq.(\ref{eq:sphaleron_energy_density_terms}). In other words, when $\xi \rightarrow 0$
\begin{align} \label{eq:boundary_1}
    f_3(1-f)=0,\quad f(2-f)-f_3=0,\quad 1-f_0=0.
\end{align}
when $\xi \rightarrow \infty$, the condition (ii) requires all terms except including $1/\xi^2$ vanish
\begin{align} \label{eq:boundary_2}
    f^\prime = f_3^\prime=f_0^\prime=\phi^\prime=0,\quad 1-f=0,\quad f_0-f_3=0.
\end{align}
Combining Eqs.~(\ref{eq:boundary_1},\ref{eq:boundary_2}), we have:
\begin{equation}
 \begin{aligned} \label{eq:sphaleron_boundary_conditions}
&\xi\rightarrow0,\quad f=f_3=\phi=0,\ f_0=1\\
&\xi\rightarrow \infty,\quad f=f_3=f_0=\phi=1.
\end{aligned}   
\end{equation}
Requiring $f$ and $f_3$ have identical boundary conditions (since both $f$ and $f_3$ characterize the $SU(2)$ gauge field), one can verify that all other combinations lead to a trivial solution or that a stable numerical solution doesn't exist. We present the sphaleron numerical solutions under boundary conditions in Eq.(\ref{eq:sphaleron_boundary_conditions}) in Fig.~\ref{fig:SM_different_profile} for the pure Standard Model scenario.

\section{Electroweak septuplet Extension to the SM: Model analysis} 
\label{sec:EWPT_model_analysis}
In this section, we will analyze the scalar septuplet extension to the SM under different EWPT scenarios, using the formalism outlined in Sec.~\ref{sec::theory_formalism}. 
As a prelude, let us review the motivation for focusing on the scalar septuplet. 
In general, for an electroweak multiplet having isospin $J$, $J$ cannot be arbitrarily large. When $J\geq 5$, the scale at which the gauge coupling Landau pole occurs would decrease to around $\Lambda_{\text{Landau}}\leq 10 $ TeV \cite{AbdusSalam:2013eya}. Furthermore, the partial wave unitarity condition for tree-level scattering amplitude constrains $J\leq 7/2$ for a complex scalar multiplet and $J \leq 4$ for a real scalar multiplet \cite{Hally:2012pu,Earl:2013jsa}. We further focus on the possibility that the neutral component, having $J_3+Y=0$, contributes to the dark matter relic density. In order to avoid stringent experimental dark matter  direct detection constraints, we require that the neutral field does not couple to $Z$ boson, which requires $Y=0$. Since only a multiplet with integer $J$ can have a $J_3=0$ component, we will focus on this scenario. Such electroweak multiplet with vanishing vev can be a dark matter candidate \cite{Chao:2018xwz,Cirelli:2005uq}.
Thus, the highest dimension for an electroweak multiplet satisfying the unitary condition and providing a viable dark matter candidate is the septuplet with $J=3$ \cite{Cirelli:2005uq}. In addition, being a member in the minimal dark matter scenario \cite{Cirelli:2005uq}, the neutral septuplet can be a viable DM candidate without imposing a $Z_2$ symmetry by hand. Depending on the various patterns of EWSB in the septuplet extension, the monopole and the sphaleron solutions can each emerge at different points in cosmic history. We, thus, compute the corresponding sphaleron energy and monopole mass. 

As discussed in the introduction, we consider three patterns of EWSB, as shown in Fig.~\ref{fig:different_steps_EWPT}. Fig.~\ref{fig:different_steps_EWPT} (a) displays the one-step EWPT to pure Higgs phase, where the additional scalar can change the sphaleron energy of the Higgs phases through thermal loops. In principle, the thermal loop corrections should also be included when analyzing patterns (b) or (c), since the EWSB occurs at hot early universe. The three dimensional effective field theory (3dEFT) is a powerful analytic method to organize the thermal corrections \cite{Kajantie:1995dw, Braaten:1995cm, Farakos:1994kx}. There are recent applications of 3dEFT to the nucleation rate computation \cite{Lofgren:2021ogg,Hirvonen:2021zej}, whose results show that the thermal correction would bias the zero temperature four dimensional model parameters (including vev) to some extent. However, zero temperature analysis can still provide a useful baseline for subsequent $T>0$ analyses.
In our present work, we mainly aim to provide a methodology for applying the sphaleron and monopole formalism to different EWSB patterns, so the zero temperature analysis is a useful starting point. When non-zero temperature effects are included, a similar analysis strategy can be applied to the thermal potential.

For our current zero-temperature analysis, we are more interested in cases (b) and (c). We label the vevs of the scalar potential stationary points in Fig.~\ref{fig:different_steps_EWPT} as, $X(v,0),\ W(0,v_\phi)$ and $Z(v_z,v_{z\phi})$. In general, $v_z\neq v$ and $v_{z\phi} \neq v_\phi$. Furthermore, when we parameterize the scalar fields and perform a model analysis, we usually regard the field vevs as input parameters. Thus, for patterns (b) and (c), we cannot use one single model analysis strategy, since the required input vevs and model parameter relationships may differ in different EWSB patterns. We will show two analysis strategies separately after the introduction of the model.

\subsection{The Model}

The general potential of the SM Higgs $H$ and another $\text{SU}(2)$ multiplet $\Phi$ can be written as \cite{Chao:2018xwz}
\begin{equation} \label{eq::potential_higgs_with_multiplet}
\begin{aligned}
V=&M_A^2\left(\Phi^{\dagger} \Phi\right)+\left\{M_B^2(\Phi \Phi)_0+\text { h.c. }\right\} \\
& -\mu^2 H^{\dagger} H+\lambda\left(H^{\dagger} H\right)^2+\lambda_1\left(H^{\dagger} H\right)\left(\Phi^{\dagger} \Phi\right) \\
&+\lambda_2\left((\overline{H} H)_1(\overline{\Phi} \Phi)_1\right)_0+\left[\lambda_3(\overline{H} H)_0(\Phi \Phi)_0+\text { h.c. }\right] \\
&+V_{\text{self}}(\Phi,\overline{\Phi}),
\end{aligned}
\end{equation}
with
\begin{equation}
    \begin{aligned}
        V_{\text{self}}(\Phi,\overline{\Phi}) =& \sum_{J=0}^{2 J} \kappa_k\left((\Phi \Phi)_k(\overline{\Phi}\,\overline{\Phi})_k\right)_0 \\
        &+\sum_{k=0}^{2 J}\left\{\kappa_k^{\prime}\left((\Phi \Phi)_k(\Phi \Phi)_k\right)_0 \right. \\
        & \left. +\kappa_k^{\prime \prime}\left((\overline{\Phi} \Phi)_k(\Phi \Phi)_k\right)_0+\text { h.c. }\right\}.
    \end{aligned}
\end{equation}
where $J$ is the multiplet isospin index, and $J=3$ is the septuplet case. The scalar multiplet self-interaction potential $ V_{\text{self}}(\Phi,\overline{\Phi})$ may be important in solving the core-cusp problem \cite{deBlok:2009sp,Tulin:2017ara}. The $\overline{H}$ and $\overline{\Phi}$ are the complex conjugate representation of $H$ and $\Phi$. As pointed out in \cite{Chao:2018xwz}, the terms $(\Phi\Phi)_1$, $(\Phi\Phi)_3$ and $(\Phi\Phi)_5$ vanish due to the property of Clebsch-Gordan coefficients. Therefore, for the self interaction potential, only terms with $k\in[0,2,4,6]$ have non-zero contributions. Further more, only terms with $k=0,2$ are independent for our septuplet example \cite{Chao:2018xwz,Cao:2022ocg}, which simplifies our model analysis.

\subsection{One-step EWPT to the mixed phase} \label{sec:one_step_EWPT_to_mixed}

In this pattern, we parameterize the general complex Higgs field ($H$), septuplet field ($\Phi$) and their complex conjugate representation ($\overline{H},\ \overline{\Phi}$) as
\begin{equation} \label{eq::Higgs_one_step}
H=\left(
\begin{array}{c}
\omega^+ \\
 \frac{1}{\sqrt{2}}(v_z+h+i \pi) \\
\end{array}
\right);\overline{H}=\left(
\begin{array}{c}
 \frac{1}{\sqrt{2}}(v_{z}+h-i \pi) \\
 -\omega^- \\
\end{array}
\right),
\end{equation}

\begin{equation} \label{eq::Septuplet_one_step}
\Phi =\left(
\begin{array}{c}
 \phi _{3,3} \\
 \phi _{3,2} \\
 \phi _{3,1} \\
 \frac{1}{\sqrt{2}}(v_{z\phi}+\phi+i \pi_\phi) \\
 \phi _{3,-1} \\
 \phi _{3,-2} \\
 \phi _{3,-3} \\
\end{array}
\right);\overline{\Phi }=\left(
\begin{array}{c}
\phi _{3,-3}^* \\
 -\phi _{3,-2}^* \\
 \phi _{3,-1}^* \\
 \frac{-1}{\sqrt{2}}(v_{z\phi}+\phi-i \pi_\phi) \\
 \phi _{3,1}^* \\
 -\phi _{3,2}^* \\
 \phi _{3,3}^* \\
\end{array}
\right).
\end{equation}
where $v_z$ and $v_{z\phi}$ are vevs of the Higgs field and septuplet field, respectively. We use the subscript $z$ to denote the vevs in the one-step EWPT (point $Z$ of Fig.~\ref{fig:different_steps_EWPT}). While for the two-step EWPT analysis in Sec.~\ref{sec:two_step_ewpt}, the vevs at point $X$ and $W$ of Fig.~\ref{fig:different_steps_EWPT} are denoted as $v$ and $v_\phi$, respectively. 

We put the septuplet vev into its neutral component.
Additional constraints need to be applied if we put the Higgs and septuplet's vevs both into real neutral components. This can be fulfilled by requiring that all the fluctuation fields (inside Higgs or septuplet) have non-negative mass-squared eigenvalues. Of course, one important constraint follows from the tadpole condition
\begin{equation} \label{eq:tadpole_constraint_fopt}
    \frac{\partial V}{\partial x_i}\bigg\vert_{\forall x_i=0}=0\ \ ,
\end{equation}
where $x_i\in [h,\pi,\omega^{\pm},\phi,\pi_\phi,\phi_{3,j},\phi_{3,j}^*]$; $j$ denotes the various subscripts that appear in $\Phi$; and  $\forall x_i=0$ means set all the field fluctuations equal to zero after the partial derivative. Subsequently, we can obtain five parameter constraints
\begin{equation}
    \begin{aligned} \label{eq::one_step_tadpole_condition}
        & \text{Im}(M_B^2) = 0, \\
        & \text{Im}(\lambda_3)=0, \\
        & \text{Im}(\kappa_0^{\prime \prime})-2\text{Im}(\kappa_0^\prime)+\frac{4[\text{Im}(\kappa_2^{\prime \prime})-2\text{Im}(\kappa_2^\prime)]}{3\sqrt{5}}=0, \\
        & \mu ^2= \lambda  v_z^2+\lambda_{13}v_{z\phi}^{2}, \\
        & M_{A}^{2}-\frac{2}{\sqrt{7}}\text{Re}(M_{B}^{2})=-\lambda_{s}v_{z\phi}^{2}-\lambda_{13}v_z^{2},
    \end{aligned}
\end{equation}
where the first three constraints arise from a single tadpole condition, which we will explain now. The tadpole condition $\partial V/\partial \pi =0$ gives us:
\begin{align} \label{eq:single-tadpole}
    c_1 \text{term}_1 + c_2 \text{term}_2 + c_3 \text{term}_3 = 0 \ \ ,
\end{align}
where $c_1,\ c_2$ and $c_3$ are constants depending on the vev of the Higgs field; $\text{term}_1$, $\text{term}_2$ and $\text{term}_3$ read
\begin{equation}
\begin{aligned}
    \text{term}_1&=\text{Im}(M_B^2),\ \ \text{term}_2=\text{Im}(\lambda_3) \ , \\
    \text{term}_3&=\text{Im}(\kappa_0^{\prime \prime})-2\text{Im}(\kappa_0^\prime)+\frac{4[\text{Im}(\kappa_2^{\prime \prime})-2\text{Im}(\kappa_2^\prime)]}{3\sqrt{5}}.
\end{aligned}
\end{equation}
Thus, we have decomposed the single constraint in Eq.(\ref{eq:single-tadpole}) into three separate constraints in Eq.(\ref{eq::one_step_tadpole_condition}), as to eliminate the mixing between $h$ and $\pi_\phi$ and simplify our analysis. 
In addition, $\lambda_{13}$ and $\lambda_s$ are two combined parameters
\begin{equation} \label{eq::lam13}
    \lambda_{13}=\frac{1}{2}\lambda_1-\frac{1}{\sqrt{14}}\lambda_3,
\end{equation}
and
\begin{equation} \label{eq::lams}
\begin{aligned}
\lambda_{s}=&\frac{1}{7}[\kappa_{0}+2\text{Re}(\kappa_{0}^{\prime})-2\text{Re}(\kappa_{0}^{\prime \prime})] \\
&+\frac{4}{21\sqrt{5}}[\kappa_{2}+2\text{Re}(\kappa_{2}^{\prime})-2\text{Re}(\kappa_{2}^{\prime \prime})]. \\
\end{aligned}
\end{equation}
As discussed in Ref.~\cite{Chao:2018xwz}, $\lambda_{13}$ enters the DM annihilation and direct detection rates, while $\lambda_s$ characterizes DM self-interactions.

The mass terms in total potential can be expressed as a series addition of mass-squared matrices:
\begin{equation} \label{eq::one_step_mass_series}
\begin{aligned}
V(H,\Phi)\supset&\frac{1}{2}\left(\begin{array}{cc} h & \phi \end{array}\right) \text{H}_{2\times 2}  \left(\begin{array}{c} h \\ \phi \end{array}\right) + \frac{1}{2}\left(\begin{array}{cc}  \pi & \pi_\phi \end{array}\right) \text{Pi}_{2\times 2}  \left(\begin{array}{c} \pi \\ \pi_\phi \end{array}\right)  \\
& + \left(\begin{array}{ccc} \omega^+ & \phi_{3,1} & \phi_{3,-1}^{*} \end{array}\right) \text{C1}_{3\times 3}\left(\begin{array}{c} \omega^- \\ \phi_{3,1}^{*}  \\ \phi_{3,-1} \end{array}\right) \\
& + \left(\begin{array}{cc}  \phi_{3,2} & \phi_{3,-2}^{*} \end{array}\right) \text{C2}_{2\times 2}\left(\begin{array}{c} \phi_{3,2}^{*}  \\ \phi_{3,-2} \end{array}\right) \\
&+\left(\begin{array}{cc}  \phi_{3,3} & \phi_{3,-3}^{*} \end{array}\right) \text{C3}_{2\times 2}\left(\begin{array}{c} \phi_{3,3}^{*}  \\ \phi_{3,-3} \end{array}\right).
\end{aligned}
\end{equation}
where we put the explicit expressions of mass-squared matrices into Appendix \ref{app:mass_matrices}. As expected, we  observe a massless pseudo-scalar particle and a massless charged Higgs particle after the computation of matrix eigenvalues. The matrix $\text{Pi}_{2\times 2}$ has one non-zero eigenvalue and $\text{C1}_{3\times 3}$ has two non-zero eigenvalues.

Let us now enumerate the constraints that we need to apply. If this number plus the quantity of input parameters is less than or equal to the degrees of freedom of model parameters, and all the parameters are involved in these constraints, we are free to move on. On the one hand, non-negative mass-squared matrix eigenvalues require at most 9 constraints: 1 from $\text{Pi}_{2\times 2}$, 2 from each other four matrices. Also, we have 5 tadpole constraints, so we have 14 parameter constraints. On the other hand, we have 19 degrees of freedom from the the model Eq.~\eqref{eq::potential_higgs_with_multiplet} (note that the complex parameter contributes 2 degrees of freedom and we need to count the SM two parameters, since the SM constraints are counted). So in total we can set at least 5 independent input parameters for this model. In principle, we might take these 5 input parameters to be $v_z,v_{z\phi},\lambda,\lambda_{13}$ and $\lambda_s$, as they will appear in our later potential analysis. However, since the Higgs field mixes with the septuplet field, it is more convenient to use the mass eigenstate and mixing angle as input parameters. The two mass eigenstates $h_1$ and $h_2$ are given by 
\begin{equation}
    \left(\begin{array}{c}
     h_1  \\
     h_2 
\end{array}\right)=\left(
\begin{array}{cc}
 \cos\theta &  \sin\theta \\
-\sin\theta & \cos\theta \\
\end{array}
\right)\left(\begin{array}{c}
     h  \\
     \phi 
\end{array}\right).
\end{equation}
with the mass-squared eigenvalues $m_1^2$ and $m_2^2$. We assign $h_2$ as SM-like Higgs particle with $m_2=125.25$ GeV. The $\lambda,\lambda_{13}$, and $\lambda_s$ can be expressed in terms of $\cos\theta,m_1^2$, and $m_2^2$. The relationships between these quantities are provided in Appendix \ref{app:mass_matrices}. 

According to our analysis in Sec.\ref{sec:classification_of_topological_field_solutions}, the formed topological field solution in the one-step EWPT to the mixed-vev phase [scenario (c)] is the sphaleron, whose energy will be computed below. According to the sphaleron scalar field configuration Eq.~\eqref{eq:multiplet_sphaleron_confg_KL_scalar}, we need to set all the fluctuation fields in Eq.~\eqref{eq::Higgs_one_step} and Eq.~\eqref{eq::Septuplet_one_step} equal to zero. Making the following replacement
\begin{equation}
    v_z\rightarrow h[\xi]v_z,\ v_{z\phi} \rightarrow \phi[\xi]v_{z\phi},
\end{equation}
we can obtain the final potential formula in one-step EWPT as
\begin{equation}
\begin{aligned} \label{eq::potential_onestep_high}
V_{\text{One}}(\xi,\mu=\frac{\pi}{2})=
&\frac{1}{2} v_{z\phi}^2 \phi[\xi]{}^2 \left[ \lambda _{13} v_z^2 h(\xi){}^2- \left(v_{z\phi}^2 \lambda _s+\lambda _{13} v_z^2\right)\right]\\
&+\frac{1}{4} v_z^2 h[\xi]{}^2 \left[\lambda  v_z^2 h(\xi){}^2-2 \left(\lambda  v_z^2+\lambda _{13} v_{z\phi}^2\right)\right] \\
&+\frac{1}{4} v_{z\phi}^4 \lambda _s \phi[\xi]{}^4 ,
\end{aligned}
\end{equation}
where $V_{\text{One}}$ represents the one-step EWPT to the mixed phase. The ground state potential reads
\begin{equation} \label{eq::potential_onestep_ground}
    V_{\text{One, ground}}=-\frac{1}{4} \left(v_{z\phi}^4 \lambda _s+\lambda  v_z^4+2 \lambda _{13} v_{z\phi}^2 v_z^2\right) .
\end{equation}

Thus far, we have finished the last task needed to solve the EOM and compute the sphaleron energy. Equations \eqref{eq::potential_onestep_high} and \eqref{eq::potential_onestep_ground} multipled by the normalization factor $\frac{\xi^2}{g^2 \Omega^4}$ constitute the potential that appears in Eq.~\eqref{eq::sphaleron_energy}. However, for the potential term $V[h(\xi),\phi(\xi)]$ that appears in the EOMs Eq.~\eqref{eq::EOMs}, we should directly use Eq.~\eqref{eq::potential_onestep_high} without any such normalization factors. 

\subsection{Two-step EWPT}
\label{sec:two_step_ewpt}

First, we expand the Higgs and septuplet fields around their extremal
scalar field configuration
\begin{equation} \label{eq::field_two_step_false_vacuum}
H=\frac{h}{\sqrt{2}}\left( \begin{array}{c} 0 \\ 1 \end{array} \right), \quad \Phi=\frac{\phi}{\sqrt{2}}\left( \begin{array}{c} 0 \\ 0 \\ 0 \\ 1 \\ 0 \\ 0 \\ 0  \end{array} \right),
\end{equation}
Then, substituting Eq.~\eqref{eq::field_two_step_false_vacuum} into Eq.~\eqref{eq::potential_higgs_with_multiplet}, we can obtain a general potential expression $V_{\text{general}}$. Secondly, applying the tadpole criteria
\begin{equation}
    \frac{\partial V_{\text{general}}}{\partial h}=\frac{\partial V_{\text{general}}}{\partial \phi}=0,
\end{equation}
we can obtain nine extremal points, which have a $\mathbb{Z}_{2}$ symmetry. These nine extremal points can be shown by mirroring Fig.~\ref{fig:different_steps_EWPT} (c) to all four quadrants. The value and hessian determinant of $X, W$, and $Z$ point in Fig.~\ref{fig:different_steps_EWPT} (c) are summarized in Table \ref{table:two_step_parameter_table}. The vev relations and defined parameters in Table \ref{table:two_step_parameter_table} read
\begin{equation} \label{eq::two_step_tadpole_condition}
\begin{aligned}
   & v=\frac{\mu }{\sqrt{\lambda }}, \quad v_\phi=\frac{\sqrt{2 \sqrt{7} \text{M}_{B}^{2}-7 \text{M}_{A}^{2}}}{\sqrt{7\lambda _s}}, \\
   & v_z^{2}=\frac{\lambda _s \left(\lambda _{13} v_\phi^2 - \lambda  v^2\right)}{\lambda _{13}^2-\lambda  \lambda _s},\\
   &v_{z\phi}^{2}=\frac{\lambda (\lambda _{13} v^2 - v_\phi^2 \lambda _s)}{\lambda _{13}^2 - \lambda  \lambda _s}, \\
  & V_z=v_\phi^4 \lambda _s+\lambda  v^4-2 \lambda _{13} v_\phi^2 v^2,
\end{aligned}
\end{equation}
where $v$ and $v_\phi$ denote vevs at points $X$ and $W$ of Fig.~\ref{fig:different_steps_EWPT} in two-step EWPT; while $v_z$ and $v_{z\phi}$ denote vevs of point $Z$ in one-step EWPT; where the definition of $\lambda_{13}$ and $\lambda_s$ are same with Eq.~\eqref{eq::lam13} and Eq.~\eqref{eq::lams}. We notice that the relationship between vevs and model parameters are different from the one-step EWPT to mixed phase, as shown in Eq.~\eqref{eq::one_step_tadpole_condition}. One can verify that, inside Eq.~\eqref{eq::two_step_tadpole_condition}, if we put the expression of $v$ and $v_\phi$ (the first line) into $v_z^2$ (the second line) and $v_{z\phi}^2$ (the third line) definitions, the $v_z^2$ and $v_{z\phi}^2$ have following relation
\begin{equation}
    \begin{aligned}
        &\mu^2=\lambda v_z^2 +\lambda_{13} v_{z\phi}^2, \\
        &M_A^2-\frac{2\text{Re}(M_B^2)}{\sqrt{7}}=-\lambda_s v_{z\phi}^2-\lambda_{13} v_z^2 .
    \end{aligned}
\end{equation}
These are just the last two relations in Eq.~\eqref{eq::one_step_tadpole_condition}, so the two analysis methods are consistent with each other. 

Let us elaborate further on the mass matrix in the two-step EWPT. The computational methods should be quite parallel with one-step scenario, where we need to start from the general field parameterization Eq.~\eqref{eq::Higgs_one_step} and Eq.~\eqref{eq::Septuplet_one_step}. While, the difference comes from the relationship between vevs and model parameters. Therefore, we can obtain the various mass matrices as displayed in Eq.~\eqref{eq::one_step_mass_series}, but with different parameter relationships. 

Returning to our potential analysis, we can express the potential as
\begin{equation}
\begin{aligned} \label{eq:general_potential_of_two_step_EWPT}    V_{\text{general}}&=\frac{1}{4} \big[\phi^2 \left(2 h^2 \lambda _{13}-2 v_\phi^ 2 \lambda _s\right)  \\
& \qquad +h^2 \left(h^2 \lambda -2 \lambda  v^2\right)+\phi^4 \lambda _s\big].
\end{aligned}
\end{equation}

Now we comment about the topological field solutions in the two-step EWPT. According to our analysis in Sec.\ref{sec:classification_of_topological_field_solutions} and our $Y=0$ septuplet setup, the topological solution at point $W$ of Fig.\ref{fig:different_steps_EWPT} (c) is the monopole, while the final electroweak vacuum gives rise to the sphaleron. We convert these two situations into different values of the vevs:
\begin{itemize}
    \item $W:$ $\text{Higgs vev}=0$, $\text{Septuplet vev}\neq 0$,  monopole
    \item $X:$ $\text{Higgs vev}\neq0$, $\text{Septuplet vev}= 0$, sphaleron
\end{itemize}
In the following, the Higgs vev and Septuplet vev are denoted as $v$ and $v_\phi$, respectively.

For purposes of deriving and solving the EOM and computing the sphaleron energy or monopole mass, we need to make the substitution $h\rightarrow h[\xi]v, \phi\rightarrow \phi[\xi]v_\phi$. Then the potential reads 
\begin{equation}
\begin{aligned} \label{eq::two_step_potential_high}
    V_{\text{Two}}(\xi,\mu=\frac{\pi}{2})=&\frac{1}{4}h[\xi]^2 v^{2} (h[\xi]^2 v^{2} \lambda -2 \lambda  v^2)+\frac{1}{4}\phi[\xi]^4 v_\phi^{4} \lambda _s \\
    &+\frac{1}{2}\phi[\xi]^2 v_\phi^{2} ( h[\xi]^2 v^{2} \lambda _{13}- v_\phi^2 \lambda _s).
\end{aligned}
\end{equation}
where $V_{\text{Two}}$ represents potential in two-step EWPT scenario, which has the identical property with Eq.~\eqref{eq::potential_onestep_high} in the sphaleron energy computation. 

The vacuum potential in two-step EWPT reads
\begin{equation} \label{eq::potential_twostep_ground}
    V_{\text{Two, ground}}=-\frac{1}{4} \left(v_\phi^4 \lambda _s+\lambda  v^4-2 \lambda _{13} v_\phi^2 v^2\right) .
\end{equation}
The specific potential expressions at point $W$ or $X$ of Fig.\ref{fig:different_steps_EWPT} (c) can be obtained from  Eq.~(\ref{eq::two_step_potential_high}) and Eq.~(\ref{eq::potential_twostep_ground}) with the simplification rules that we listed in the paragraph below Eq.~(\ref{eq:general_potential_of_two_step_EWPT}).

\begin{figure}[t]
\center 
\includegraphics[width=7cm]{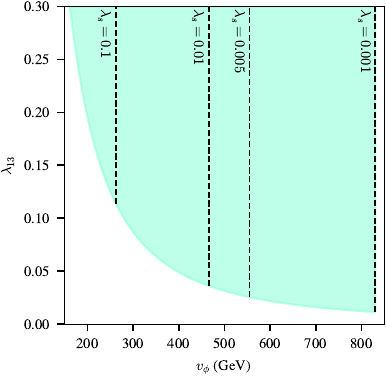}
\caption{Parameter ranges that satisfy the two-step EWPT. The intersection of aquamarine color region and left hand side of $\lambda_s$ vertical dashed line represents the feasible parameter region. The larger value of $\lambda_s$, the smaller viable parameter region.
}
\label{fig::parameter_two_step}
\end{figure}

To fulfill a two-step EWPT, additional parameter constraints should be applied. As shown in Fig.~\ref{fig:different_steps_EWPT} (c), we require that our universe undergoes the transition pattern: $O \rightarrow W \rightarrow X$. In principle, such constraints should be considered in a thermal bath with thermal corrections included to the effective potential. Our zero-$T$ analysis here only serves as an estimation. Here are the requirements
\begin{itemize}
\item[1.] $O$ must be a secondary local minimum, this requires 
\begin{equation}
    \lambda _s>0,
\end{equation}
\item[2.] $V(W)>V(X)$, this implies
\begin{equation} \label{eq::VW>VX}
\lambda _s v_\phi^{4}< \lambda v^{4},
\end{equation}
\item[3.] The hessian determinant $\text{Hess}(X)>0$, this requires
\begin{equation}  \label{eq::HessX>0}
\lambda _{13} v^2-v_\phi^2 \lambda _s>0,
\end{equation}
\item[4.] The hessian determinant $\text{Hess}(W)>0$, this implies
\begin{equation} \label{eq:HessW>0}
\lambda _{13} v_\phi^2-\lambda  v^2>0,
\end{equation}

\item[5.] if we require the point $Z$ exist, we need to solve the equations:
\begin{equation}
\begin{aligned}
    &v_{z}^{2}=\frac{\lambda _s \left(\lambda _{13} v_\phi^2 - \lambda  v^2\right)}{\lambda _{13}^2-\lambda  \lambda _s}, \\ &v_{z\phi}^{2}=\frac{\lambda (\lambda _{13} v^2 - v_\phi^2 \lambda _s)}{\lambda _{13}^2 - \lambda  \lambda _s},
\end{aligned}
\end{equation}
with the constraints Eq.~\eqref{eq::HessX>0} and Eq.~\eqref{eq:HessW>0}, the conditions $v_{z}^{2}>0$ and $v_{z\phi}^{2}>0$ require
\begin{equation}
\lambda _{13}^2-\lambda  \lambda _s>0\ \ .
\end{equation}
one would observe that the hessian determinant $\text{Hess}(Z)<0$ under all above criteria, so the mixed point is not a stationary point.
\end{itemize}
These constraints are not totally independent, since the constraint Eq.~\eqref{eq::HessX>0} can be derived out from Eq.~\eqref{eq::VW>VX} and Eq.~\eqref{eq:HessW>0}, and the latter two conditions are of crucial importance. Overall, we have five input parameters: $v,v_\phi,\lambda, \lambda_s$ and $\lambda_{13}$. The parameter ranges that satisfy the two-step EWPT are shown in Fig.~\ref{fig::parameter_two_step}. In this plot, the lower bound of the aquamarine region is constrained by Eq.~\eqref{eq::HessX>0}. The vertical bound refer to the lines for constant $\lambda_s$, which gives the maximum value bound of $v_\phi$ according to Eq.~(\ref{eq::VW>VX}). The smaller the value of $\lambda_s$, the larger unconstrained parameter region we would have. At the end of the first step, constrained by the effective portal coupling, the septuplet vev cannot be arbitrarily small.

\begin{table}
\caption{Parameter table for the two-step EWPT. In two-step EWPT, it is advantageous to express all quantities in terms of $v$ and $v_\phi$. Accordingly, both the potential in Eq.~(\ref{eq:general_potential_of_two_step_EWPT}) and the associated Hessian determinant have been expressed in terms of these two vevs.}
\label{table:two_step_parameter_table}
\centering
\begin{tabular}{||c c c c c||} 
 \hline
& $h$ & $\phi$ & $V$  & Hessian Determinant \\ [0.5ex] 
 \hline\hline
 $O$ & 0 & 0 & 0 & $\lambda \lambda_{s} v^{2} v_\phi^{2}$\\[2pt]
 $X$ & $v$ & 0 & $-\frac{1}{4} \lambda  v^4$  & $2 \lambda  v^2 \left(\lambda _{13} v^2-v_\phi^2 \lambda _s\right)$\\[2pt]
 $W$ &0 & $v_\phi$ & $-\frac{1}{4} \lambda _s v_\phi^4 $ & $2 v_\phi^2 \lambda _s \left(\lambda _{13} v_\phi^2-\lambda  v^2\right)$\\[3pt]
 $Z$ & $v_{z}$ & $v_{z\phi}$ & $\frac{\lambda  \lambda _s V_z}{4 \left(\lambda _{13}^2-\lambda  \lambda _s\right)}$ & $\frac{4 \lambda  \lambda _s \left(\lambda  v^2-\lambda _{13} v_\phi^2\right) \left(\lambda _{13} v^2-v_\phi^2 \lambda _s\right)}{\lambda _{13}^2-\lambda  \lambda _s}$\\[2pt]
 \hline
\end{tabular}
\label{table:1}
\end{table}

\section{Sphaleron energy and monopole mass in different EWPT scenarios}\label{sec:sphaleron_energy}
The formal sphaleron energy and monopole mass can be defined as \cite{Quiros:1999jp}
\begin{equation} \label{eq::sphaleron_B_definition}
E_{\text{sph}}/m_\text{mon}=B_{\text{sph/mon}}\cdot \frac{4\pi \Omega}{g},
\end{equation}
where $\Omega=246.22$ GeV is the zero temperature Higgs vev and $g$ is the weak coupling constant. The value of $B_{\text{sph}}$ and $B_{\text{mon}}$ is defined as the radial integral of Eq.~\eqref{eq::sphaleron_energy} and Eq.~\eqref{eq::monopole_mass_explicit_expression}, respectively. In the SM, where the EWPT is shown in pattern (a) in Fig.~\ref{fig:different_steps_EWPT}, the sphaleron $B_{\text{sph}}=1.900506$. We will compute the value of the $B_{\text{sph}}$ in Fig.\ref{fig:different_steps_EWPT} (b) at point $Z$ and the value of $B_{\text{mon}}$ in Fig.\ref{fig:different_steps_EWPT} (c) at point $W$ in this section. 
\begin{figure*}[t]
\center
\includegraphics[width=7cm]{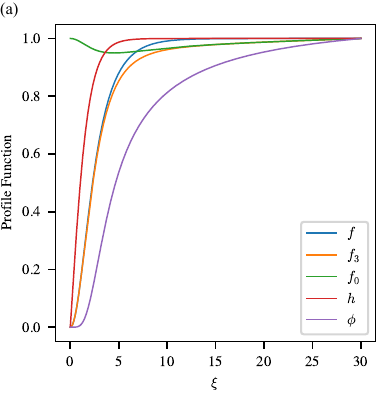}
\hspace{0.01cm}
\includegraphics[width=7cm]{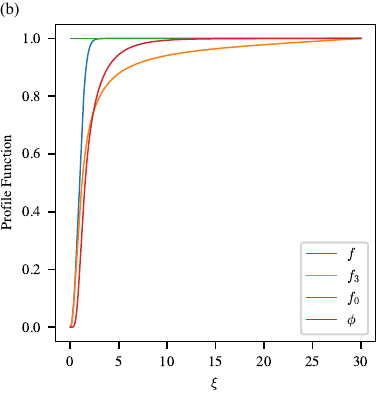}
\caption{The radial profile functions in different EWSB patterns. Figure (a) denotes the EWPT to the mixed phase, where we take $v_\phi=1$ GeV, $\lambda_{13}=0.05$ and $\lambda_s=0.005$. In this scenario, both the Higgs field $h$ and multiplet field $\phi$ obtain vev, and the sphaleron energy value $B_{\text{sph}}=1.900535$. Figure (b) represents the two-step EWPT (at point $W$ of pattern (c) in Fig.~\ref{fig:different_steps_EWPT}), where we take $v=0$ GeV, $v_\phi=500$ GeV, $\lambda_{13}=0.05$ and $\lambda_s=0.005$. At this stage, the Higgs field's profile function doesn't appear due to its vanishing vev, and the value of monopole mass in this stage is $B_{\text{mon}}=5.001145$.
}
\label{fig:One_and_Two_EWPT_profile_function}
\end{figure*}

\begin{figure}[h]
\center
\includegraphics[width=8cm]{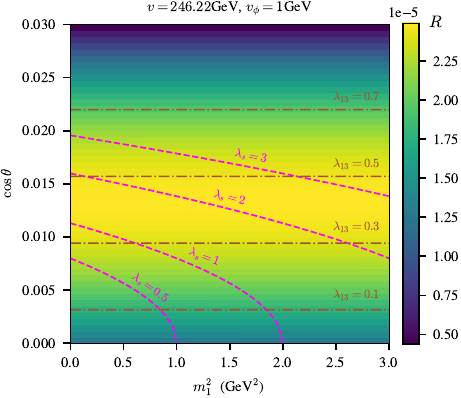}
\caption{Sphaleron energy in one-step EWPT to mixed phase. The model free parameters are chosen as $v_z,v_{z\phi},m_2,\cos\theta,$ and $m_1$. These parameters represent the Higgs vev ($v_z=246.22$ GeV), the septuplet vev ($v_{z\phi}=1$ GeV), the Higgs mass ($m_2=125.25$ GeV), the mixing angle ($\cos\theta$), and the light scalar mass eigenvalue $m_1$, where the last two parameters are regarded as scan input parameters. The color bar value denotes the relative change of sphaleron energy to the SM scenario, which is defined as $R=(E_{\text{Mixed}}- E_{\text{SM}})/E_{\text{SM}}$, where $E_{\text{Mixed}}$ and $E_{\text{SM}}$ are the sphaleron energy in the mixed phase and SM scenarios, respectively. The SM sphaleron energy has a $B_\text{sph}$ value of $1.900506$, where $B_\text{sph}$ is defined in Eq.~\eqref{eq::sphaleron_B_definition}. Note that there is a multiplication factor of $10^{-5}$ in the color bar. The dashed line and dot-dashed line represent the contour lines of $\lambda_s$ and $\lambda_{13}$, respectively. The contour line of $\lambda$ is not plotted, as its absolute maximum relative change with respect to the SM value is smaller than $|-4\times 10^{-4}|$. The SM scalar self-interaction coefficient is $\lambda=0.129383$.
}
\label{fig:one_step_sphaleron_energy}
\end{figure}
Before presenting the results for $E_{\text{sph}}$ and $m_\text{mon}$, we first comment on the constraints of the septuplet model. For one-step EWPT to the mixed phase, we consider the $\rho$ parameter constraint, which imposes significant limit on the septuplet vev. In the two-step EWPT scenario, since the neutral septuplet field can contribute to the dark matter relic density, we focus primarily on constraints from the dark matter direct detection experiments. Additionally, there are constraints from the collider experiments, such as the $h\rightarrow \gamma\gamma$ channel and the electroweak precision tests (the oblique $S$, $T$, and $U$ parameters). A detailed analysis of these constraints would diverge from the main focus of this work, and we plan to conduct a comprehensive phenomenological study in future research.

\subsection{One-Step EWPT to the Mixed Phase}
In this situation, both Higgs field and septuplet field obtain vev after the phase transition, while the $v_\phi$ should be constrained by the $\rho$ parameter. The $\rho$ parameter under multiple electroweak scalars is defined as
\begin{equation}
\rho=\sum_{i}\frac{[J_{i}(J_{i}+1)-Y_{i}^{2}] v_{i}^{2}}{2Y_{i}^{2}v_{i}^{2}},
\end{equation}
where $J_i$ is the total isospin, $Y_i$ denotes the hypercharge. In our situation, we have two scalar fields, one is the Higgs field with $J=\frac{1}{2}$ and $Y=\frac{1}{2}$, another is the additional multiplet with a general $J$ and $Y=0$. Then, the $\rho$ parameter is given by
\begin{equation}
\rho=1+2 J (J+1)\frac{v_\phi^{2}}{v^{2}},
\end{equation}
the larger isospin of the multiplet, the stronger constraint are imposed on $v_\phi$. The value of $\rho$ parameter is analyzed in Ref.~\cite{ParticleDataGroup:2024cfk} with $\rho=1.00031 \pm 0.00019$. Within $95\%$ significance level, $v_\phi$ is constrained to
\begin{equation}
    v_\phi^2 \lesssim \frac{20.648}{J(J+1)} \text{GeV}^2.
\end{equation}
 For our septuplet case ($J=3$), we are safe to take $v_\phi=1$ GeV.

The computation of sphaleron energy can be separated into two parts: (i) obtain the solutions of field profile function from the EOMs in Eq.~\eqref{eq::EOMs}; (ii) put the solutions into the sphaleron energy expression of Eq.~\eqref{eq::sphaleron_energy}. For the first step, we present the solutions of the field's profile function in Fig.~\ref{fig:One_and_Two_EWPT_profile_function} (a), with the choice of parameters $\lambda_{13}=0.05$ and $\lambda_s=0.005$. Under this parameter choice, the $B_{\text{sph}}=1.900535$, which is quite close to the SM $B_\text{sph}$ value. Further more, we perform a parameter scan of $B_\text{sph}$ and present the result in Fig.~\ref{fig:one_step_sphaleron_energy}, where the input parameters are chosen according to the analysis of Sec.\ref{sec:one_step_EWPT_to_mixed}. Since the Higgs vev is overwhelmingly larger than the septuplet vev, the sphaleron energy differs little from pure SM case. Moreover, the light scalar mass has almost no impact on the sphaleron energy, while as the mixing angle parameter $\cos\theta$ increases, the sphaleron energy initially increases before subsequently decreasing. In summary, in the one-step EWPT scenario, if we only consider one scalar multiplet extension, constrained by the $\rho$ parameter, the additional multiplet has negligible influence on the SM sphaleron energy. 

We would like to make some comments about Georgi-Machacek model \cite{Georgi:1985nv} where the vevs for more than one additional new mutliplets can be large, while satisfy the $\rho$ parameter constraint. The formalism to analyze this case will be the same as discussed here, but then including one additional field vev. We might anticipate a significantly different result for the sphaleron energy in this case. We defer a detailed study to future work.

\subsection{Two-step EWPT}
\begin{figure}[h]
\center
\includegraphics[width=7.5cm]{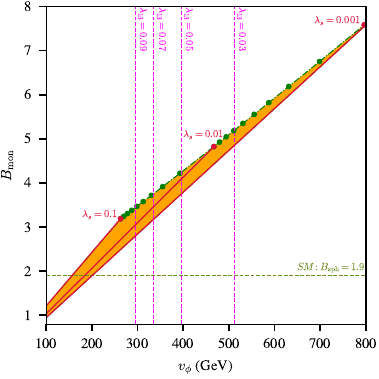}
\caption{Monopole mass $B_{\text{mon}}$ in the first step (point $W$) of two-step EWPT. The intersection of the orange region and right hand side of the dashed vertical $\lambda_{13}$ line represents the viable monopole mass region. The bottom olive drab dashed line denotes the SM sphaleron $B_{\text{sph}}$ value, which is approximately equal to 1.9. The orange region represents the Monopole mass $B_{\text{mon}}$ under a feasible choice of $v_\phi$ and $\lambda_s$. The bottom red line represents $\lambda_s=0.001$, while the upper red line denotes $\lambda_s=0.1$; $\lambda_s$ continuously increases counter clockwise; the red or green dots represent the truncated points limited by the $v_\phi$ in Eq.~\eqref{eq::VW>VX}. The constraint from $\lambda_{13}$ in Eq.~\eqref{eq:HessW>0} rejects the left hand side of the vertical $\lambda_{13}$ line. }
\label{fig:two_step_sphaleron_energy}
\end{figure}

As shown in Fig.~\ref{fig:different_steps_EWPT} (c), the first step is $C1:\ O\rightarrow W$ and the second step is $C2:\ W\rightarrow X$. The vev of the multiplet at point $W$ is unconstrained, since the $\rho$ parameter is measured at point $X$ in zero temperature universe. Thus, the monopole mass at point $W$ can reach a sizable value, and we can anticipate that $B_\text{mon}$ could differ significantly from the SM value of $B_\text{sph}$. Parallel to the one-step EWPT analysis, we show the solutions of the field profile function in the right panel of Fig.~\ref{fig:One_and_Two_EWPT_profile_function}. Also, we perform a parameter scan over $B_\text{mon}$ and present the result in Fig.~\ref{fig:two_step_sphaleron_energy}.

In Fig.~\ref{fig:two_step_sphaleron_energy}, the intersection between the orange region and the right hand side of the vertical dashed $\lambda_{13}$ line represents the unconstrained monopole mass domain. From Eq.~\eqref{eq::two_step_potential_high}, we observe that the portal effective coupling $\lambda_{13}$ does not affect the potential term $V_{\text{Two}}$ under the vanishing Higgs vev ($v=0$) scenario, so that $\lambda_{13}$ doesn't alter the monopole mass at $W$. On the other hand, the greater value of $\lambda_s$, the higher value of the monopole mass. Therefore, the influence of $\lambda_{13}$ and $\lambda_s$ to $B_\text{mon}$ at two-step EWPT differs from their influence on $B_\text{sph}$ in one-step case. This difference can be deduced from the different potential configuration in one-step case with Eq.~\eqref{eq::potential_onestep_high} and two-step case with Eq.~\eqref{eq::two_step_potential_high}. 

It is interesting to observe that there is a sizeable orange region with $B_{\text{mon}}$ greater than the SM value of $B_{\text{sph}}$. If this pattern persists at $T>0$; if our universe undergoes a first order EWPT during the first step ($C1$); and if there exists sufficient BSM CPV to create the baryon asymmetry, this asymmetry can be well preserved at point $W$.
For demonstration in the real triplet extension, see Refs.~\cite{Patel:2012pi,Blinov:2015sna,Inoue:2015pza,Niemi:2020hto}.
In general, the second step $C2$ to the Higgs phase could either preserve or erase this baryon asymmetry. If the second step is first order and if the sphaleron energy at point $X$ is sufficiently large, then this asymmetry can be preserved in the final Higgs phase. A complete analysis of this possibility for the $T>0$ general electroweak multiplet case will appear in a future study.

Finally, we comment on model constraints implied by dark matter (DM) phenomenology. The CP-even or CP-odd neutral septuplet fields can serve as the dark matter candidate. (See Appendix~\ref{app::mass_matrices_two_step_EWPT} for the mass matrices and analyses.)  
The work in \cite{Chao:2018xwz} analyzes the corresponding dark matter relic density and direct detection constraints. For the latter, the DM-nucleus spin independent cross section can be written as $\sigma_{\text{SI}}\propto g^2_{\text{eff}}$, with
\begin{equation}
    g_{\text{eff}}=f_N \frac{2\lambda_{13}}{m_h^2}+\frac{3}{4}f_Tf_N^{\text{PDF}}.
\end{equation}
where $f_N \approx 0.287$ for proton; $f_N^{\text{PDF}}=0.526$ denotes the second momentum of nucleon parton distribution function (PDF); $f_T$ is the coefficient function of twist-two quark bilinear operator \cite{Chao:2018xwz}; and $m_h$ represents the Higgs particle mass. The first term in $g_{\text{eff}}$ is the dominant contribution of $\sigma_{\text{SI}}$. 
In our parameter scan, we take $\lambda_{13}$ smaller than $0.1$ both in Fig.~\ref{fig:one_step_sphaleron_energy} and Fig.~\ref{fig:two_step_sphaleron_energy}. For dark matter mass in $10^2-10^4$ GeV, we have confirmed that our parameter choices with $\lambda_{13}=0.03$ are unlimited by the newest dark matter direct search constraint \cite{PandaX:2024qfu,2024arXiv241017036A} .

\section{Conclusion} \label{sec:conclusion}

Determining the origin of the cosmic baryon asymmetry remains an important challenge at the interface of particle and nuclear physics with cosmology. Among various possible baryogenesis mechanisms, we focus on electroweak baryogenesis (EWBG), which naturally connects with the Higgs mechanism. 
In this context, we perform a detailed study of the formalism for baryon number dynamics -- a key EWBG ingredient -- in the context of a first order EWPT catalyzed by the extension of the SM with a scalar electroweak multiplet.
 Furthermore, we choose the complex septuplet as a case of study to compute the sphaleron energy and monopole mass. \par

For the theory formalism, we first classify the topological field solutions. For two-step electroweak symmetry breaking, wherein the new multiplet vacuum expectation value (vev) alone breaks electroweak symmetry during the first step, representations having non-zero or zero hypercharge admit sphaleron or monopole solutions, respectively. When a non-vanishing Higgs doublet vev contributes to the symmetry breaking, either in the second step or in a one-step symmetry breaking scenario, only sphaleron solutions emerge. Experimental dark matter direct detection constraints imply that only zero hypercharge multiplets with vanishing vevs in the present broken symmetry phase may contribute to the dark matter relic density. Such multiplets would, thus, be associated with monopole-catalyzed baryon number violation during the penultimate era of electroweak symmetry breaking. Then, we discuss other formal features of sphaleron and monopole field solutions. 
For the sphaleron induced BNV, we show that two previous configurations, Manton-Klinkhamer (MK) and Klinkhamer-Laterveer (KL) are equivalent up to a gauge transformation. Furthermore, we construct the general dimensional sphaleron and monopole matrices, define the general sphaleron energy and monopole mass expressions and discuss the choices of boundary conditions for field equations of motion.

For the sphaleron or monopole computation, we first analyse the multiplet extension model's parameter constraint in one-step EWPT to mixed phase and two-step EWPT scenario separately. In both scenarios, we have five potential related independent parameters, the Higgs and septuplet vev, the Higgs and septuplet effective self couplings, and the Higgs-septuplet effective portal coupling. In the scenario of a single-step EWPT leading to a mixed-vev phase, we find that the additional multiplet's contribution to the sphaleron energy is negligible, primarily due to the constraint imposed by the electroweak $\rho$ parameter. Conversely, in a two-step EWPT scenario, the monopole mass can achieve sufficiently large values so as to suppress the monopole density and allow for preservation of baryon asymmetry generated during this step, assuming a first order EWPT and the requisite CP-violation. For the latter case, we analyze the constraints on model parameters from the dark matter phenomenology. On the other hand, the DM self-interaction coupling, $\lambda_s$, is less constrained from the direct search experiments. We find that the monopole mass exhibits complementary dependencies on these two parameters: $\lambda_{13}$ does not directly affect the value of the monopole mass, while the monopole mass increases monotonically with $\lambda_s$.

In the future, numerous studies can be conducted based on this work. For instance, the computation of sphaleron energy under thermal corrections and the computation of one-step EWPT with the Georgi-Machacek model; the computation of the next-to-leading-order BNV rate; the accurate estimation of the monopole density, etc.

\begin{acknowledgments}

We thank Wei Chao, Tobias Diez, Yucheng Qiu,  Xu-Xiang Li, Nicholas Manton, Tudor Stefan RATIU, Tuomas V. I. Tenkanen, Bingrong Yu, Michael Zantedeschi and Jiang Zhu for useful discussions during this work. We thank the anonymous referee for providing insightful feedback. M.J. Ramsey-Musolf, Y. Wu, and W. Zhang were supported in part by the National Natural Science Foundation of China under grant no. 12375094 and 11975150 and by the Ministry of Science and Technology of China under grant no. WQ20183100522. M. J. Ramsey-Musolf also gratefully acknowledges support under the Double First Class Plan of the Shanghai Jiao Tong University and sponsorship from Shanghai Tang Junyuan Education Foundation. W. Zhang is supported by the National Natural Science Foundation of China under grant No.12405120, Start-up Funds for Young Talents of Hebei University (No.521100224226).
\end{acknowledgments}

\newpage

\appendix \label{app:whole_appendix}

\section{Relationship between chern-simons number and $\mathcal{P}(t_a,t_b)$} \label{app:Relationship between chern-simons number and Ptatb}
In the main text, Eq.(\ref{eq:Ptatb_definition}), we define an integral $\mathcal{P}(t_a,t_b)$ read
\begin{equation}
\begin{aligned} \label{eq:Ptatb_definition_appendix}
	\mathcal{P}(t_a,t_b)&\equiv\int_{t_a}^{t_b} dt \int_{-\infty}^{+\infty} d^3 x\ \partial_\mu K^\mu \\
	&=\int d^{3} x K^{0}\bigg\vert^{t=t_a}_{t=t_b}+\int^{t_b}_{t_a}dt \int_{S} \vec{K}\cdot \vec{dS}.
\end{aligned}
\end{equation} 
from which we defined the winging number and sphaleron baryonic charge, as shown in Eq.(\ref{eq:winding_number_definition}) and Eq.(\ref{eq:sphaleron_baryonic_charge_definition}), respectively. Now we want to connect $\mathcal{P}(t_a,t_b)$ into the definition of chern-simons number. 

The chern-simons number $N_{\text{CS}}$ is defined as:
\begin{equation} \label{eq:chern_simons_definition}
	N_{\text{CS}}(t=t_b)-N_{\text{CS}}(t=t_a)=\mathcal{P}(t_a,t_b),
\end{equation}
we can take $N_{\text{CS}}(t=-\infty)=0$ for the vacum configuration. If we further assign $t_b=t_0$, one have
\begin{equation} 
\begin{aligned} \label{eq:Ncs_t0} 
		N_{\text{CS}}(t=t_0)&=Q_B=\mathcal{P}(-\infty,t_0) \\
	&=\int d^{3} x K^{0}\bigg\vert^{t=t_{0}}_{t=-\infty}+\int^{t_{0}}_{-\infty}dt \int_{S} \vec{K}\cdot \vec{dS} ,
\end{aligned}
\end{equation}
Suppose the gauge field $A_i^a$ falls off faster than $1/r$, the second term in the second line of Eq.~(\ref{eq:Ncs_t0}) equals zero. So that,
\begin{equation} \label{eq:NCS_Q_B_K0_only}
\begin{aligned}
		&N_{\text{CS}}(t=t_0) = \int d^{3} x K^{0}\bigg\vert_{t=t_0} \\
		&=\frac{g^{2}}{16\pi^{2}} \int d^{3}x \epsilon^{ijk} \left((\partial_{i}A_{j}^{a})A_{k}^{a}+\frac{g}{3}f^{abc}A_{i}^{a}A_{j}^{b}A_{k}^{c} \right).
\end{aligned}
\end{equation}

Define $\bm{\mathcal{{A}}}_{i}\equiv i g A_{i}^{a}\frac{\sigma^a}{2}$, where $\sigma^a$ being the Pauli matrix. One can verify that the following relation holds
\eqal{1}{ \label{eq:trace_gauge_term}
&\text{Tr}\left( (\partial_{i}\bm{\mathcal{{A}}}_{j})\bm{\mathcal{{A}}}_{k}-\frac{1}{3}\bm{\mathcal{{A}}}_{i}[\bm{\mathcal{{A}}}_{j},\bm{\mathcal{{A}}}_{k}] \right) \\
&=-\frac{g^{2}}{2}\left( (\partial_{i}A_{j}^{a})A_{k}^{a} +\frac{g}{3} f^{abc} A_{i}^{a}A_{j}^{b}A_{k}^{c} \right),
}
so that $N_{\text{CS}}(t=t_{0})$ read
\begin{equation}
\begin{aligned}
&N_{\text{CS}}(t=t_{0})=\\
&\quad -\frac{1}{8\pi^{2}} \int d^{3}x \epsilon^{ijk} \text{Tr}\left( (\partial_{i}\bm{\mathcal{{A}}}_{j})\bm{\mathcal{{A}}}_{k}-\frac{1}{3}\bm{\mathcal{{A}}}_{i}[\bm{\mathcal{{A}}}_{j},\bm{\mathcal{{A}}}_{k}] \right),
\end{aligned}
\end{equation}
Under a gauge transformation, $\bm{\mathcal{{A}}^\prime}_{i} =U \bm{\mathcal{{A}}}_{i} U^{-1}+(\partial_{i} U)U^{-1}$, we have
\begin{equation}
	\begin{aligned}
&N(\bm{\mathcal{{A^{\prime}}}},t=t_{0}) 
	 =N(\bm{\mathcal{{A}}},t=t_{0})\\
  &\quad +\frac{1}{8\pi^{2}}\int d^{3}x \epsilon^{ijk}\partial_{i}\text{Tr}(U\bm{\mathcal{{A}}}_{j}(\partial_{k}U^{-1})) \\
	& \quad +\frac{1}{24\pi^{2}} \int d^{3}x \epsilon^{ijk} \text{Tr} (U(\partial_{i}U^{-1})U(\partial_{j}U^{-1})U(\partial_{k}U^{-1})).
	\end{aligned}
\end{equation}
where the second term is a total derivative, which will vanish as long as the gauge field vanish at the boundaries. The last term is important and has the topological meaning, which characterizes the number of times the $SU(2)$ is covered by a three-sphere.

\section{Gauge choice with $A_i^a$ falls off faster than $1/r$} \label{app:gauge_choice_Aia_falls_faster_1over_r}
We comment on the gauge transformation matrix $U_c$ used in MK configuration \cite{Klinkhamer:1984di} which causes the gauge field $A_i^a$ to decay more rapidly than $1/r$. We start from the gauge field configuration when $\mu=\pi/2$ (with an additional rotation in Ref.~\cite{Klinkhamer:1984di}):
\begin{equation}
	A_i^a=-\frac{2f(\xi)}{gr^2}\epsilon_{iab}x_b,
\end{equation}
applying the gauge transformation:
\begin{equation} \label{eq:Uc_definition}
U_c=\exp((-1)^n i\Theta(r) \vec{T}\cdot \hat{\vec{x}}),
\end{equation}
where $n=1$ or $2$ regarding on the convention of sign; where $\Theta(r)$ runs rapidly from $0$ to $\pi$ when $r$ runs from $0$ to $\infty$; 
where $\vec{T}$ are the SU(2) generators, and $\hat{\vec{x}}=\vec{x}/r$. The gauge field configuration is transformed into
\begin{equation} \label{eq:gauge_field_confg_falls_faster_1overr}
\begin{aligned}
		A_i^a&=\frac{(1-2f(\xi))\cos\Theta(r)-1}{gr^2}\epsilon_{iab}x_b \\
		&\quad +(-1)^n \frac{(1-2f(\xi))\sin\Theta(r)}{gr^3}(\delta_{ia} r^2 - x_i x_a) \\
		&\quad +\frac{(-1)^n}{g}\frac{d\Theta(r)}{dr}\frac{x_i x_a}{r^2},
\end{aligned}
\end{equation}
where the resulting $A_i^a$ falls faster than $1/r$.  

We compute the sphaleron baryonic charge based on the gauge field configuration Eq.~(\ref{eq:gauge_field_confg_falls_faster_1overr}) with a general $n$ values:
\begin{equation}
\begin{aligned}
Q_{B}&=N_{\text{CS}}(t=t_0)\\
&=\int d^{3} x K^{0}\bigg\vert_{t=t_{0}}=\begin{cases}
\frac{1}{2} & \text{when } n=1 \\
-\frac{1}{2} & \text{when } n=2
\end{cases}.    
\end{aligned}
\end{equation}
Note that, there is a typo in Ref.\cite{Klinkhamer:1984di}, where they obtained $Q_B=\frac12$ with $n=2$ in their definition of Eq.(\ref{eq:Uc_definition}).

\section{Other sphaleron configurations}\label{app:other_configurations}
\subsection{AKY configuration}
Under the general spherically symmetric ansatz, the gauge field configuration is written as \cite{Akiba:1988ay}
\begin{equation}
\begin{aligned}
        A_j^a(x)&=\frac{1}{g}\big[D(r)\epsilon_{jam}x_m +B(r) (r^2 \delta_{ja}-x_j x_a)\\
    &\quad + C(r) x_j x_a \big],
\end{aligned}
\end{equation}
The Higgs field is written as
\begin{equation}
    H(x)=\frac{v}{\sqrt{2}}\left[H(r)+iK(r) \frac{\vec{\sigma}\cdot \vec{\hat{r}}}{2} \right] \left(\begin{array}{c}
         0  \\
         1 
    \end{array} \right).
\end{equation}
where $D(r),B(r), C(r), H(r)$ and $K(r)$
are radial functions. Usually, the radial gauge condition sets $C(r)=0$. A recent systematic investigation on Akiba-Kikuchi-Yanagida (AKY) configuration is presented in Ref.~\cite{Matchev:2025ivr}.

\subsection{KKB configuration}
Start form a set of orthonormal vectors \cite{Kleihaus:1991ks}
\begin{equation}
\begin{aligned}
& \mathbf{u}_1(\phi)=(\cos \phi, \sin \phi, 0), \\
& \mathbf{u}_2(\phi)=(0,0,1) ,\\
& \mathbf{u}_3(\phi)=(\sin \phi,-\cos \phi, 0) ,
\end{aligned}
\end{equation}
The fields are expanded as follows
\begin{equation}
\begin{aligned}
& A_i^a(\mathbf{r})=u_j^i(\phi) u_k^a(\phi) w_j^k(\rho, z), \\
& a_i(\mathbf{r})=u_j^i(\phi) a_j(\rho, z), \\
& H(\mathbf{r})=\tau^i u_j^i(\phi) h_j(\rho, z) \frac{v}{\sqrt{2}}\left(\begin{array}{l}
0 \\
1
\end{array}\right) .
\end{aligned}
\end{equation}
where we've changed the field labels to make them consistent with the convention in this study.

\section{Boltzmann equation of monopole fermion scattering}
\label{app:Boltzmann_eq_of_monopole_fermion}
Consider a particle species $a$ undergoing the scattering process
\begin{align}
    a+X\rightarrow Y \ ,
\end{align}
where $X$ and $Y$ denote multi-particle states. Assuming that the final state $Y$ is in thermal equilibrium, the Boltzmann equation for species $a$ is given by \cite{Luty:1992un}
\begin{align}
    \dot{n}_a+3Hn_a =-\langle \sigma v\rangle (n_a n_X-n_a^{\rm eq}n_X^{\rm eq}) \ ,
\end{align}
where $n_a$ is the number density of $a$ (with $n_a^{\rm eq}$ its equilibrium value), the dot denotes the time derivative and $H$ is the Hubble constant. The thermal average of the scattering cross section is defined as
\begin{equation}
\begin{aligned}
\langle \sigma v\rangle&=\sum_{aX\leftrightarrow Y}\int d\pi_a d\pi_X d \pi_Y (2\pi)^4 \delta^4(p_a+p_X-p_Y)\\
&\quad \quad \times |\mathcal{A}(aX\rightarrow Y)|^2 \frac{f_a^{\rm eq} f_X^{\rm eq}}{n_a^{\rm eq} n_X^{\rm eq}} \ \ \ ,
 \end{aligned}
\end{equation}
where $d\pi_X$ and $d \pi_Y$ represent the phase space integrals as in Ref.~\cite{Luty:1992un}. Here, $\mathcal{A}$ denotes the transition amplitude, and we have neglected the CP violation, such that
\begin{align}
    |\mathcal{A}(aX\rightarrow Y)|^2=|\mathcal{A}(Y\rightarrow a X)|^2.
\end{align}

We now extend the analysis to baryon washout induced by the BNV scattering between monopoles and fermions. The Boltzmann equations for the baryon and anti-baryon are given by \cite{Ellis:1982bz,Brennan:2024sth}
\begin{align}
\label{eq:baryon_boltzmann_eq}
\dot{n}_B+3Hn_B&=-\langle \sigma_{\Delta B\neq 0} v\rangle_f g_f (n_fn_M-n_f^{\rm eq}n_M^{\rm eq}), \\
\label{eq:anti_baryon_boltzmann_eq}
\dot{n}_{\bar{B}}+3Hn_{\bar{B}}&=-\langle \sigma_{\Delta B\neq 0} v\rangle_{\bar{f}} g_f (n_{\bar{f}} n_{\bar{M}} - n_{f}^{\rm eq} n_{M}^{\rm eq}), 
\end{align}
where $g_f$ is the number of fermion degrees of freedom; $n_B$, $n_f$ and $n_M$ denote the number density of baryons, fermions and monopoles, respectively; while $n_{\bar{B}}$, $n_{\bar{f}}$ and $n_{\bar{M}}$ denote those of anti-baryons, anti-fermions and anti-monopoles. We assume that (i) the CP violating effect in the monopole-fermion scattering process is negligible, $\langle \sigma v\rangle_f=\langle \sigma v\rangle_{\bar{f}}$; (ii) monopoles are in thermal equilibrium, i.e., $n_M=n_{\bar{M}}=n_M^{\rm eq}$. Defining the net baryon and net fermion densities as
\begin{align}
    \bar{n}_B&\equiv n_B-n_{\bar{B}}\ ,\\
    \bar{n}_f&\equiv n_f-n_{\bar{f}} \   \ .
\end{align}
From Eq.(\ref{eq:baryon_boltzmann_eq}) and Eq.(\ref{eq:anti_baryon_boltzmann_eq}), we obtain the Boltzmann equation for the net baryon number density
\begin{equation}
    \dot{\bar{n}}_B+3H\bar{n}_B=-\langle \sigma_{\Delta B\neq 0} v\rangle_fg_f\bar{n}_fn_M \ \ .
\end{equation}

\section{Sphaleron Energy Computation}\label{app:Esph}
In this appendix, we provide the computation details of the sphaleron energy for the Yang-Mills term and the covariant derivative term.
\subsection{Yang-Mills term}
Under a general \text{SU}(2) representation, the Yang-Mills term transforms
\begin{equation}
\begin{aligned}
F^{aij}F^{a}_{ij}
&=F^{aij}F^{b}_{ij}\frac{1}{2S(R)}{\rm Tr}[\{T^{a},T^{b}\}] \\
&=\frac{1}{S(R)}{\rm Tr}[F^{aij}T^{a}\cdot F^{b}_{ij}T^{b}],
\end{aligned}
\end{equation}
where $S(R)$ is the Dynkin index, and we use ${\rm Tr}[\{T^{a},T^{b}\}]=2S(R)\delta^{ab}$. 
Since
\begin{equation}
F^{a}_{ij}T^{a}=\partial_{i}A^{a}_{j}T^{a}-\partial_{j}A^{a}_{i}T^{a}+g\epsilon^{abc}A_{i}^{b}A_{j}^{c}T^{a},
\end{equation}
and
\begin{equation}
\begin{aligned}
\epsilon^{abc}A_{i}^{b}A_{j}^{c}T^{a}&=\epsilon^{bca}T^{a}A_{i}^{b}A_{j}^{c},\\
&=\frac{1}{i}[T^{b},T^{c}]A_{i}^{b}A_{j}^{c}, \\
&=\frac{1}{i}[A_{i}^{b}T^{b}A_{j}^{c}T^{c}-A_{j}^{c}T^{c}A_{i}^{b}T^{b}] .
\end{aligned}
\end{equation}
where we have used the fact that $[T^{b},T^{c}]=i\epsilon^{bca}T^{a}$ for all \text{SU}(2) multiplets. We can deduce that the Yang-Mills term is invariant for different \text{SU}(2) multiplet representations. 

\subsection{Kinetic term}
For a general \text{SU}(2) multiplet, it's covariant derivative reads
\begin{equation}
(D_i\Phi)=\partial_i \Phi - igA^a_i T^a\Phi - ig^{\prime}a_iX\Phi ,
\end{equation}
Since our sphaleron construction occurs in spherical coordinates, the index $i\in [r, \theta, \phi]$. The kinetic term in the second phase of KL sphaleron configuration reads

\begin{widetext}
\begin{equation}
\begin{aligned}
(D_i\Phi)^{\dagger}(D_i\Phi)&=(\partial_i \Phi)^{\dagger}(\partial_i \Phi)+g^2 \langle\Phi^{\dagger}|J^bJ^a|\Phi \rangle A^a_i A^b_i + g^{\prime 2}\langle \Phi^{\dagger}|X^{2}|\Phi \rangle a^{i}a^{i}+2gg^{\prime}A_{i}^{3}a_{i}J^{3}X \Phi^\dagger \Phi ,\\
&=(\partial_i \Phi)^{\dagger}(\partial_i \Phi)+h^{2}g^{2}[\frac{v^{2}}{4}(J(J+1)-(J^{3})^{2})A_{\mu}^{+}A^{\mu-}+\frac{v^{2}}{2}(J^{3})^{2}A_{\mu}^{3}A^{\mu 3}]\\ 
&\ \ \ \ +g^{\prime 2}(J^{3})^{2}h^{2}\frac{v^{2}}{2}(a_{r}^{2}+a_{\theta}^{2}+a_{\phi}^{2}) 
-g g^{\prime} (J^{3})^2 v^2 h^2 (\frac{a_{\theta} A_{\theta}^{3}}{r^2}+\frac{a_{\theta} A_{\phi}^{3}}{(r \sin \theta )^2})\ ,
\end{aligned}
\end{equation}
\end{widetext}
where
\begin{equation}
\begin{aligned}
A_{\mu}^{+}A^{\mu-} &= \frac{(A_{\theta}^{1})^2+(A_{\theta}^{2})^2}{r^2}+\frac{(A_{\phi}^{1})^2+(A_{\phi}^{2})^2}{r^2 {\rm sin}^{2}\theta} ,\\
A_{\mu}^{3}A^{\mu 3}&=\frac{(A_{\theta}^{3})^{2}}{r^{2}}+\frac{(A_{\phi}^{3})^{2}}{r^{2}{\rm sin}^{2}\theta}\ \ .
\end{aligned}
\end{equation}
where we need to know the explicit expression of $A_i^a$, with $i\in[r,\theta,\phi]$ being the spherical coordinates label and $a\in[1,2,3]$ being the \text{SU}(2) generators label. The expressions of $A_i^a$ can be  computed through Eq.~\eqref{eq:multiplet_sphaleron_confg_KL_gauge}. 
\subsection{General energy form}
The U(1) field sphaleron energy computation is straightforward, so we don't list the result here. Finally, we scale the sphaleron energy in following way \cite{Quiros:1999jp}:

\begin{widetext}
\begin{equation}
\int d^3 x \left(\frac{1}{4}F_{ij}^aF_{ij}^a + \frac{1}{4}f_{ij}f_{ij}+ (D_i\Phi)^{\dagger}(D_i\Phi)\right) \rightarrow \frac{4 \pi \Omega}{g}\int d\xi \left(\frac{1}{4}F_{ij}^aF_{ij}^a(\xi) + \frac{1}{4}f_{ij}f_{ij}(\xi)+ (D_i\Phi)^{\dagger}(D_i\Phi)(\xi)\right),
\end{equation}
\end{widetext}
where we add the dimensionless radial parameter $(\xi)$ to each component to label the differences before and after the transformation. The sphaleron energy expressions read
\begin{widetext}
\begin{equation}
\begin{aligned}
\frac{1}{4}F_{ij}^{a}F_{ij}^{a}(\xi) &= \sin ^{2} \mu\left(\frac{8}{3} f^{\prime 2}+\frac{4}{3} f_{3}^{\prime 2}\right)+\frac{8}{\xi^{2}} \sin ^{4} \mu\left\{\frac{2}{3} f_{3}^{2}(1-f)^{2}+\frac{1}{3}\left\{f(2-f)-f_{3}\right\}^{2}\right\} ,\\
\frac{1}{4}f_{ij}f_{ij}(\xi)&=\frac{4}{3}\left(\frac{g}{g^{\prime}}\right)^{2}\left\{f_{0}^{\prime 2} \sin ^{2} \mu +\frac{2}{\xi^{2}} \sin ^{4} \mu\left(1-f_{0}\right)^{2}\right\} ,\\
(D_{i}\Phi)^{\dagger}(D_{i}\Phi)(\xi)&=\frac{v^{2}_\phi}{\Omega^{2}}\left\{\frac{1}{2} \xi^{2} \phi^{\prime 2}+\frac{4}{3} \phi^{2} \sin ^{2} \mu \left\{\left(J(J+1)-J_{3}^{2}\right)(1-f)^{2}+J_{3}^{2}\left(f_{0}-f_{3}\right)^{2}\right\}\right\} \ \ \ .
\end{aligned}
\end{equation}
\end{widetext}

\begin{widetext}
\section{Mass Matrices in the one-step EWPT under \text{SU}(2) Higgs plus septuplet model} \label{app:mass_matrices}
In this appendix, we list the explicit mass matrices that appear in Eq.~\eqref{eq::one_step_mass_series}. 

\subsection{Higgs Matrix}
\begin{equation}
\text{H}_{2\times 2}=\left(
\begin{array}{cc}
 2\lambda  v_z^2 &  2\lambda _{13} v_z v_{z\phi} \\
2 \lambda _{13} v_z v_{z\phi} & 2v_{z\phi}^2 \lambda_{s} \\
\end{array}
\right),
\end{equation}
where $\lambda_{13}$ and $\lambda_s$ are two combined parameters that defined in equations \eqref{eq::lam13} and \eqref{eq::lams}. The two mass eigenstates $h_1$ and $h_2$ can be obtained via a rotation
\begin{equation}
\left(\begin{array}{c}
     h_1  \\
     h_2 
\end{array}\right)=\left(
\begin{array}{cc}
 \cos\theta &  \sin\theta \\
-\sin\theta & \cos\theta \\
\end{array}
\right)\left(\begin{array}{c}
     h  \\
     \phi 
\end{array}\right) ,
\end{equation}
the mass eigenvalues for $h_1$ and $h_2$ are $m_1$ and $m_2$, respectively, which reads
\begin{equation}
    \begin{aligned}
        m_1^2&=\lambda v_z^2+\lambda_s v_{z\phi}^2-\sqrt{(\lambda v_z^2-\lambda_s v_{z\phi}^2)^2+4\lambda_{13}^2 v_z^2 v_{z\phi}^2},\\
        m_2^2&=\lambda v_z^2+\lambda_s v_{z\phi}^2+\sqrt{(\lambda v_z^2-\lambda_s v_{z\phi}^2)^2+4\lambda_{13}^2 v_z^2 v_{z\phi}^2},
    \end{aligned}
\end{equation}
we will assign $h_2$ as SM-like higgs particle, and take $m_2=125.25$ GeV. The $\lambda,\lambda_{13}$ and $\lambda_s$ can be expressed by 
\begin{equation}
    \begin{aligned}
        &\lambda=\frac{m_1^2 \cos\theta^2+m_2^2 \sin\theta^2}{2v_z^2},\\
        &\lambda_s=\frac{m_1^2 \sin\theta^2+m_2^2 \cos\theta^2}{2 v_{z\phi}^2},\\
        &\lambda_{13}=\frac{(m_2^2-m_1^2)\sin2\theta}{4v_z v_{z\phi}}\ \ .
    \end{aligned}
\end{equation}
Since $v_z=246.22$ GeV and we choose $v_{z\phi}=1$ GeV, the input parameters are $\cos\theta,m_1^2$.

\subsection{Pseudo-Scalar Matrix}
\begin{equation}
\text{Pi}_{2\times 2}=\left(
\begin{array}{cc}
 0 & 0 \\
 0 & \frac{4 \text{Re}(M_B^2)}{\sqrt{7}}+v_{z\phi}^2 \kappa _{\pi}+\frac{2 \text{Re}(\lambda_3) v_z^2}{\sqrt{14}} \\
\end{array}
\right),
\end{equation}
where 
\begin{equation}
    \kappa_{\pi}=\frac{2}{7}[\text{Re}(\kappa_0^{\prime \prime})-4\text{Re}(\kappa_0^{\prime})]+\frac{8}{21\sqrt{5}}[\text{Re}(\kappa_2^{\prime \prime})-4\text{Re}(\kappa_2^{\prime})].
\end{equation}

\subsection{Charged Higgs Matrices}

    \begin{equation}
\text{C1}_{3\times 3}=\left(
\begin{array}{ccc}
 0 & -\frac{v_z v_{z\phi} \lambda _2 }{2 \sqrt{14}} & -\frac{v_z v_{z\phi} \lambda _2 }{2 \sqrt{14}} \\
 -\frac{v_z v_{z\phi} \lambda _2 }{2 \sqrt{14}} & M_A^2+v_{z\phi}^2 \kappa _{\text{122}}+ v_z^2(\frac{\lambda_1}{2}+\frac{\lambda_2}{4\sqrt{42}}) & \frac{2 \text{Re}(M_B^2)}{\sqrt{7}}+\frac{\text{Re}(\lambda _3) v_z^2}{\sqrt{14}}-v_{z\phi}^2 \kappa _{\text{123}} \\
 -\frac{v_z v_{z\phi} \lambda _2 }{2 \sqrt{14}} & \frac{2 \text{Re}(M_B^2)}{\sqrt{7}}+\frac{\text{Re}(\lambda _3) v_z^2}{\sqrt{14}}-v_{z\phi}^2 \kappa _{\text{123}} & M_A^2+v_{z\phi}^2 \kappa _{\text{122}}+ v_z^2(\frac{\lambda_1}{2}-\frac{\lambda_2}{4\sqrt{42}})  \\
\end{array}
\right),
\end{equation}
where
\begin{equation}
    \begin{aligned}
        \kappa _{\text{122}}&=\frac{\kappa_2-4\text{Re}(\kappa_2^{\prime \prime})}{21\sqrt{5}}-\frac{\text{Re}(\kappa_0^{\prime \prime})}{7} ,\\
        \kappa_{\text{123}}&=\frac{1}{7}(\kappa_0+2\text{Re}(\kappa_0^\prime)-\text{Re}(\kappa_0^{\prime \prime}))+\frac{1}{21\sqrt{5}}(3\kappa_2+8\text{Re}(\kappa_2^\prime)-4\text{Re}(\kappa_2^{\prime \prime})) \ \ .
    \end{aligned}
\end{equation}
The three eigenvalues of matrix $\text{C1}_{3\times 3}$ are difficult to obtain. However, we can numerically calculate them, and we find that one of these eigenvalues equal to zero. This zero eigenvalue correspond to the massless charged Higgs particle.

\begin{equation}
\text{C2}_{2\times 2}=\left(
\begin{array}{cc}
 M_A^2+v_{z\phi}^2 \kappa _{\text{211}}+v_z^2(\frac{\lambda_1}{2}+\frac{\lambda_2}{2\sqrt{42}}) & -\frac{2 \text{Re}(M_B^2)}{\sqrt{7}}-\frac{\text{Re}(\lambda _3) v_z^2}{\sqrt{14}}+v_{z\phi}^2 \kappa _{\text{212}} \\
-\frac{2 \text{Re}(M_B^2)}{\sqrt{7}}-\frac{\text{Re}(\lambda _3) v_z^2}{\sqrt{14}}+v_{z\phi}^2 \kappa _{\text{212}}^* & M_A^2+v_{z\phi}^2 \kappa _{\text{211}}+v_z^2(\frac{\lambda_1}{2}-\frac{\lambda_2}{2\sqrt{42}}) \\
\end{array}
\right),
\end{equation}
where
\begin{equation}
    \begin{aligned}
        \kappa _{\text{211}}&=\frac{2\sqrt{5}}{21}[\kappa_2-\text{Re}(\kappa_2^{\prime \prime})]-\frac{\text{Re}(\kappa_0^{\prime \prime})}{7}, \\
        \kappa_{\text{212}}&=\frac{1}{7}(\kappa_0+2\kappa_0^\prime-\kappa_0^{\prime \prime})+\frac{2\sqrt{5}}{21}(2\kappa_2^\prime-\kappa_2^{\prime \prime})\ \ .
    \end{aligned}
\end{equation}

\begin{equation}
\text{C3}_{2\times 2}=\left(
\begin{array}{cc}
 M_A^2+v_{z\phi}^2 \kappa _{\text{311}}+v_z^2\left(\frac{\lambda_1}{2}+\frac{1}{4}\sqrt{\frac{3}{14}}\lambda_2\right) & \frac{2 \text{Re}(M_B^2)}{\sqrt{7}}+\frac{\text{Re}(\lambda _3) v_z^2}{\sqrt{14}}+v_{z\phi}^2 \kappa _{\text{312}} \\
\frac{2 \text{Re}(M_B^2)}{\sqrt{7}}+\frac{\text{Re}(\lambda _3) v_z^2}{\sqrt{14}}+v_{z\phi}^2 \kappa _{\text{312}}^* & M_A^2+v_{z\phi}^2 \kappa _{\text{311}}+v_z^2\left(\frac{\lambda_1}{2}-\frac{1}{4}\sqrt{\frac{3}{14}}\lambda_2\right) \\
\end{array}
\right),
\end{equation}
where
\begin{equation}
    \begin{aligned}
        \kappa _{\text{311}}&=-\frac{\text{Re}(\kappa_0^{\prime \prime})}{7}+\frac{\sqrt{5}\text{Re}(\kappa_2^{\prime \prime})}{21}, \\
        \kappa _{\text{312}}&=-\frac{1}{7}(\kappa_0+2\kappa_0^\prime-\kappa_0^{\prime \prime})+\frac{\sqrt{5}}{21}(\kappa_2+2\kappa_2^\prime-\kappa_2^{\prime \prime})\ \ .
    \end{aligned}
\end{equation}

\section{Mass matrices in the Higgs vacuum of two-step EWPT under SU(2) Higgs plus septuplet model} \label{app::mass_matrices_two_step_EWPT}
In this appendix, we present the Higgs plus septuplet model's mass matrices at the end of two-step EWPT (point $X$ of Higgs vacuum at Fig.~\ref{fig:different_steps_EWPT} (c)). The structure of mass matrices are different from Eq.~\eqref{eq::one_step_mass_series} in one-step scenario. The mass matrix part of the potential can be written as
\begin{equation} \label{eq::two_step_mass_series}
\begin{aligned}
V(H,\Phi)\supset&\frac{1}{2}\left(\begin{array}{cc} h & \phi \end{array}\right) \text{H}^\prime_{2\times 2}  \left(\begin{array}{c} h \\ \phi \end{array}\right) + \frac{1}{2}\left(\begin{array}{cc}  \pi & \pi_\phi \end{array}\right) \text{Pi}^\prime_{2\times 2}  \left(\begin{array}{c} \pi \\ \pi_\phi \end{array}\right) + \left(\begin{array}{cc}  \phi_{3,i} & \phi_{3,-i}^{*} \end{array}\right) \text{Ci}^\prime_{2\times 2}\left(\begin{array}{c} \phi_{3,i}^{*}  \\ \phi_{3,-i} \end{array}\right),
\end{aligned}
\end{equation}
where the charged Higgs field $\omega^{\pm}$ is massless, and therefore not included in the list; where the matrix for the CP-even neutral scalar fields is given as follows:
\begin{equation} \label{eq::two_step_CP_even_matrix}
\text{H}_{2\times 2}^\prime=\left(
\begin{array}{cc}
 2\lambda v^2 & 0 \\
 0 & \lambda_{13} v^2 \\
\end{array}
\right),
\end{equation}
which corresponds to the SM Higgs field mass ($2\lambda v^2$) and CP-even neutral septuplet field mass ($\lambda_{13} v^2$). The CP-odd neutral field mass matrix reads
\begin{equation} \label{eq::two_step_CP_old_matrix}
  \text{Pi}_{2\times 2}^\prime = \left(
\begin{array}{cc}
 0 & 0 \\
 0 & 2 M_A^2+\frac{1}{14} v^2 \left(7 \lambda_1+\sqrt{14} \lambda_3 \right) \\
\end{array}
\right),
\end{equation}
where only the septuplet fields obtain a non-zero mass eigenvalue. The CP-even or CP-odd neutral scalar fields of the septuplet field, as described in Eqs.~\eqref{eq::two_step_CP_even_matrix} and \eqref{eq::two_step_CP_old_matrix}, may contribute to the dark matter relic density. The mass matrix of the charged component of the septuplet field is given by
\begin{equation} \label{eq::two_step_charged_matrices}
    \text{Ci}^\prime_{2\times 2}=\left( \begin{array}{cc}
        M_A^2+\frac{1}{2}\lambda_1 v^2 + \frac{1}{4\sqrt{42}}i\lambda_2 v^2 & (-1)^{i+1}(M_A^2+\frac{\text{Re}(\lambda_3)v^2}{\sqrt{14}})  \\
        (-1)^{i+1}(M_A^2+\frac{\text{Re}(\lambda_3)v^2}{\sqrt{14}}) & M_A^2+\frac{1}{2}\lambda_1 v^2 - \frac{1}{4\sqrt{42}}i\lambda_2 v^2
    \end{array}\right).
\end{equation}
where the index $i\in[1,2,3]$ represents different charged fields in Eq.~\eqref{eq::two_step_mass_series}. The charged matrix $\text{Ci}^\prime_{2\times 2}$ has a similar structure with the septuplet charged matrix in Ref.~\cite{Chao:2018xwz}, while our relationship between $M_A^2$ and $M_B^2$ is different with Ref.~\cite{Chao:2018xwz}. The difference comes form the model tadpole conditions, which are distinct for the different EWSB patterns (for example, two-step EWPT analysis in our case).

\end{widetext}

\bibliography{ref}

\end{document}